\documentclass[a4paper,11pt]{article}
\pdfoutput=1

\usepackage{jinstpub}
\usepackage[utf8]{inputenc}
\usepackage{subcaption}
\usepackage{booktabs}

\newcommand{\ttbar}[0]{t\bar{t}}
\newcommand{\bbbar}[0]{b\bar{b}}
\newcommand{\ttH}[0]{\ttbar H}

\newcommand{\ttbb}[0]{\ttbar\!+\!\bbbar}

\newcommand{\Fig}[1]{Figure~\ref{#1}}
\newcommand{\Figs}[1]{Figures~\ref{#1}}
\newcommand{\Eqn}[1]{Equation~\ref{#1}}
\newcommand{\Apx}[1]{Appendix~\ref{#1}}

\newcommand{\bhad}[0]{b_{had}}
\newcommand{\blep}[0]{b_{lep}}
\newcommand{\qi}[0]{q_1}
\newcommand{\qii}[0]{q_2}

\newcommand{\bi}[0]{b_1}
\newcommand{\bii}[0]{b_2}

\author{M. Erdmann,}
\author{E. Geiser,}
\author{Y. Rath,}
\author{and M. Rieger}
\affiliation{III. Physics Institute A, RWTH Aachen University, Otto-Blumenthal-Str., 52074 Aachen, Germany}
\emailAdd{rieger@physik.rwth-aachen.de}

\title{\boldmath{Lorentz Boost Networks: Autonomous Physics-Inspired Feature Engineering}}
\abstract{
We present a two-stage neural network architecture that enables a fully autonomous and comprehensive characterization of collision events by exclusively exploiting the four momenta of final-state particles.
We refer to the first stage of the architecture as Lorentz Boost Network (LBN).
The LBN allows the creation of particle combinations representing rest frames.
The LBN also enables the formation of further composite particles, which are then transformed into said rest frames by Lorentz transformation. The properties of the composite, transformed particles are compiled in the form of characteristic variables that serve as input for a subsequent network.
This second network has to be configured for a specific analysis task such as the separation of signal and background events. Using the example of the classification of $\ttH$ and $\ttbb$ events, we compare the separation power of the LBN approach with that of domain-unspecific deep neural networks (DNN).
We observe leading performance with the LBN, even though we provide the DNNs with extensive additional input information beyond the particle four momenta.
Furthermore, we demonstrate that the LBN forms physically meaningful particle combinations and autonomously generates suitable characteristic variables.
}


\begin{document}

\maketitle
\flushbottom

\section{Introduction}
\label{sec:introduction}

A key element in the development of machine learning methods is the exploitation of the underlying structure of the data through appropriate architectures.
For example, convolutional neural networks make use of local, translation-invariant correlations in image-like data and compile them into characteristic features \cite{Hinton, Ciresan, ILSVRC2012, He2015, deOliveira:2015xxd, Aurisano:2016jvx, Komiske:2016rsd, Kasieczka:2017nvn, Erdmann:2017str, Shilon:2018xlp, Delaquis:2018zqi, Adams:2018bvi}.

In particle physics, characteristic features are used to identify particles, to separate signal from background processes, or to perform calibrations.
These features are usually expressed in manually engineered variables based on physics reasoning.
As input for machine learning methods in particle physics, these so-called high-level variables were in many cases superior to the direct use of particle four momenta, often referred to as low-level variables.
Recently, several working groups have shown that deep neural networks can perform better if they are trained with both specifically constructed variables and low-level variables \cite{Baldi2014:exotics, Baldi2015:higgs, Adam2015:kaggle, Guest2016:jetflavor, Baldi:2016fql, Shimmin:2017PhRvD, Louppe:2017ipp, Datta:2017rhs, Erdmann:ttH2017, Butter:2017cot, deOliveira:2017pjk, Stoye:2017, Sirunyan:2018mvw}.
This observation suggests that the networks extract additional useful information from the training data.

In this paper, we attempt to autonomize the process of finding suitable variables that describe the main characteristics of a particle physics task.
Through a training process, a neural network should learn to construct these variables from raw particle information.
This requires an architecture tailored towards the structure of particle collision events.
We will present such an architecture here and investigate its impact on the separation power for signal and background processes in comparison to domain-unspecific deep neural networks.
Furthermore, we will uncover characteristic features which are identified by the network as particularly suitable.

Particle collisions at high energies produce many particles which are often short-lived.
These short-lived particles decay into particles of the final state, sometimes through a cascade of multiple decays.
Intermediate particles can be reconstructed by using energy-momentum conservation when a parent particle decays into its daughter particles.
By assigning to each particle a four-vector defined by the particle energy and its momentum vector, the sum of the four-vectors of the daughter particles gives the four-vector of the parent particle.
For low-energy particle collisions, bubble chamber images of parents and their daughter particles were recorded in textbook quality.
Evidently, here the decay angular distributions of the daughter particles are distorted by the movement of the parent particle, but can be recovered in the rest frame of the parent particle after Lorentz transformation.

The same principles of particle cascades apply to high-energy particle collisions at colliders.
Here, the particles of interest are, for example, top quarks or Higgs bosons, which remain invisible in the detector due to their short lifetimes, but can be reconstructed from their decay products.
Exploiting the properties of such parent particles and their daughter particles in appropriate rest frames is a key to the search for high-level variables characterizing a physics process.

For the autonomous search for sensitive variables, we propose a two-stage network, composed of a so-called Lorentz Boost Network (LBN) followed by an application-specific deep neural network (NN).
The LBN takes only the four-vectors of the final-state particles as input.
In the LBN there are two ways of combining particles, one to create composite particles, the other to form appropriate rest frames.
Using Lorentz transformations, the composite particles are then boosted into the rest frames.
Thus, the decay characteristics of a parent particle can be exploited directly.

Finally, characteristic features are derived from the boosted composite particles: masses, angles, etc.
The second network stage (NN) then takes these variables as input to solve a specific problem, e.g. the separation of signal and background processes.
While the first stage constitutes a novel network architecture, the latter network is interchangeable and can be adapted depending on the analysis task.

This paper is structured as follows.
First, we explain the network architecture in detail.
Second, we present the simulated dataset we use to investigate the performance of our architecture in comparison to typical deep neural networks.
Thereafter, we review the particles, rest frames, and characteristic variables created by the network to gain insight into what is learned in the training process, before finally presenting our conclusions.

\section{Network architecture}
\label{sec:architecture}

In this section we explain the structural concept on which the Lorentz Boost Network (LBN) is based and introduce the network architecture.

The measured final state of a collision event is typically rather complex owing to the high energies of the particles involved.
In principle, all available information is encoded in the particles' four-vectors, but the comprehensive extraction of relevant properties poses a significant challenge.
To this end, physicists engineer high-level variables to decipher and factorize the probability distributions inherent to the underlying physics processes.
These high-level variables are often fed into machine learning algorithms to efficiently combine their descriptive power in the context of a specific research question.
However, the consistent result of \cite{Baldi2014:exotics, Baldi2015:higgs, Adam2015:kaggle, Guest2016:jetflavor, Baldi:2016fql, Shimmin:2017PhRvD, Louppe:2017ipp, Datta:2017rhs, Erdmann:ttH2017, Butter:2017cot, deOliveira:2017pjk, Stoye:2017, Sirunyan:2018mvw} is that the combination of both low- and high-level variables tends to provide superior performance.
This observation suggests that low-level variables potentially contain additional, useful information that is absent in hand-crafted high-level variables.

\begin{figure}[h!tbp]
    \centering
    \includegraphics[width=1.0\textwidth]{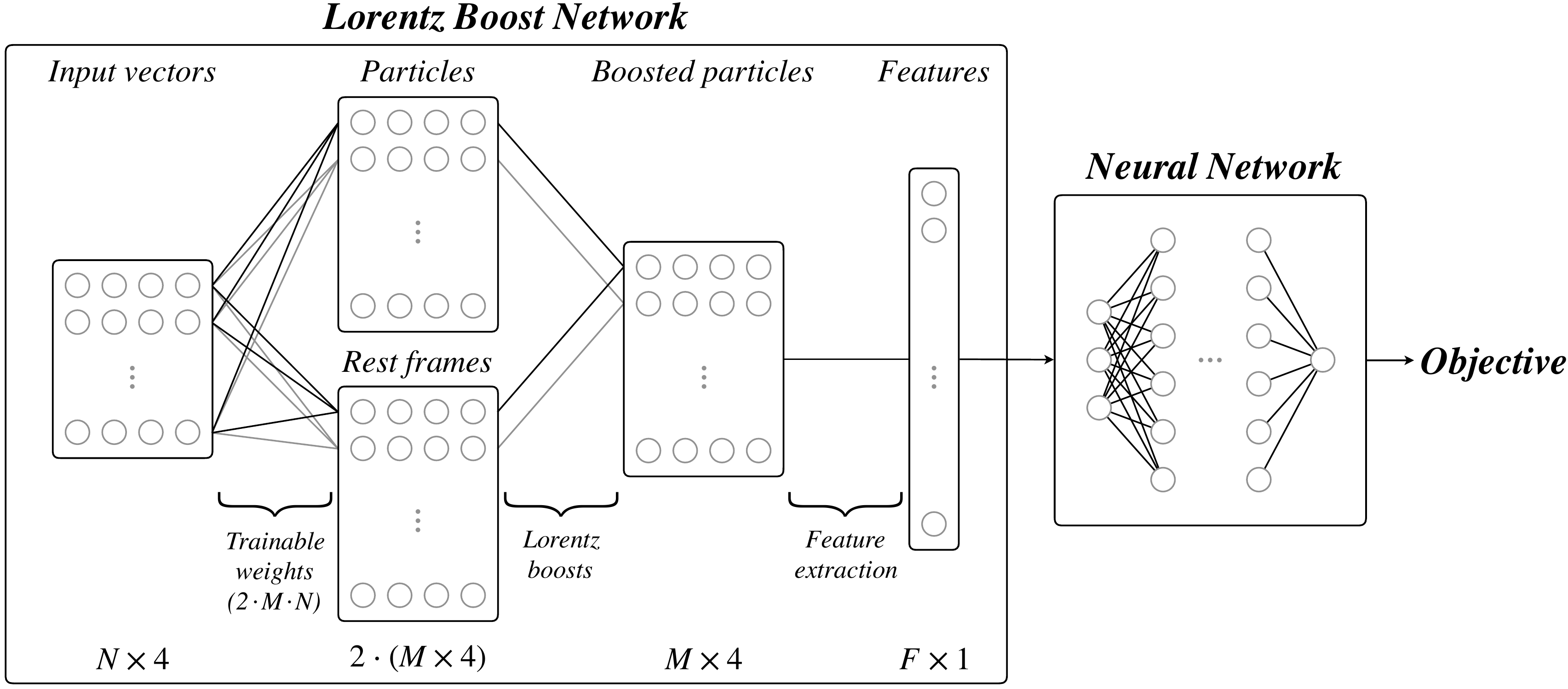}
    \caption{
        The two-stage deep neural network architecture consists of the Lorentz Boost Network (LBN) and a subsequent deep neural network (NN).
        In the LBN, the input four-vectors ($E, p_x, p_y, p_z$) are combined in two independent ways before each of the combined particles is boosted into its particular rest frame, which is formed from a different particle combination.
        The boosted particles are characterized by variables which can be, e.g., invariant masses, transverse momenta, pseudorapidities, and angular distances between them.
        These features serve as input to the second network designed to accomplish a particular analysis task.
    }
    \label{fig:lbn_arch}
\end{figure}

The aim of the LBN is, given only low-level variables as input, to autonomously determine a comprehensive set of physics-motivated variables that maximizes the relevant information for solving the physics task in the subsequent neural network application.
\Fig{fig:lbn_arch} shows the proposed two-stage network architecture (LBN+NN) in detail.
The first stage is the LBN and constitutes the novel contribution.
It consists of several parts, namely the combination of input four-vectors to particles and rest frames, subsequent Lorentz transformations, and the extraction of suitable high-level variables.
The second stage can be some form of deep neural network (NN) with an objective function depending on the specific research question.

The LBN combines $N$ input four-vectors, consisting of energies $E$ and momentum components $p_x, p_y, p_z$, to create $M$ particles and $M$ corresponding rest frames according to weighted sums using trainable weights.
Through Lorentz transformation, each combined particle is boosted into its dedicated rest frame.
After that, a generic set of features is extracted from the properties of these $M$ boosted particles.

\begin{figure}[h!tbp]
    \centering
    \includegraphics[width=0.7\textwidth]{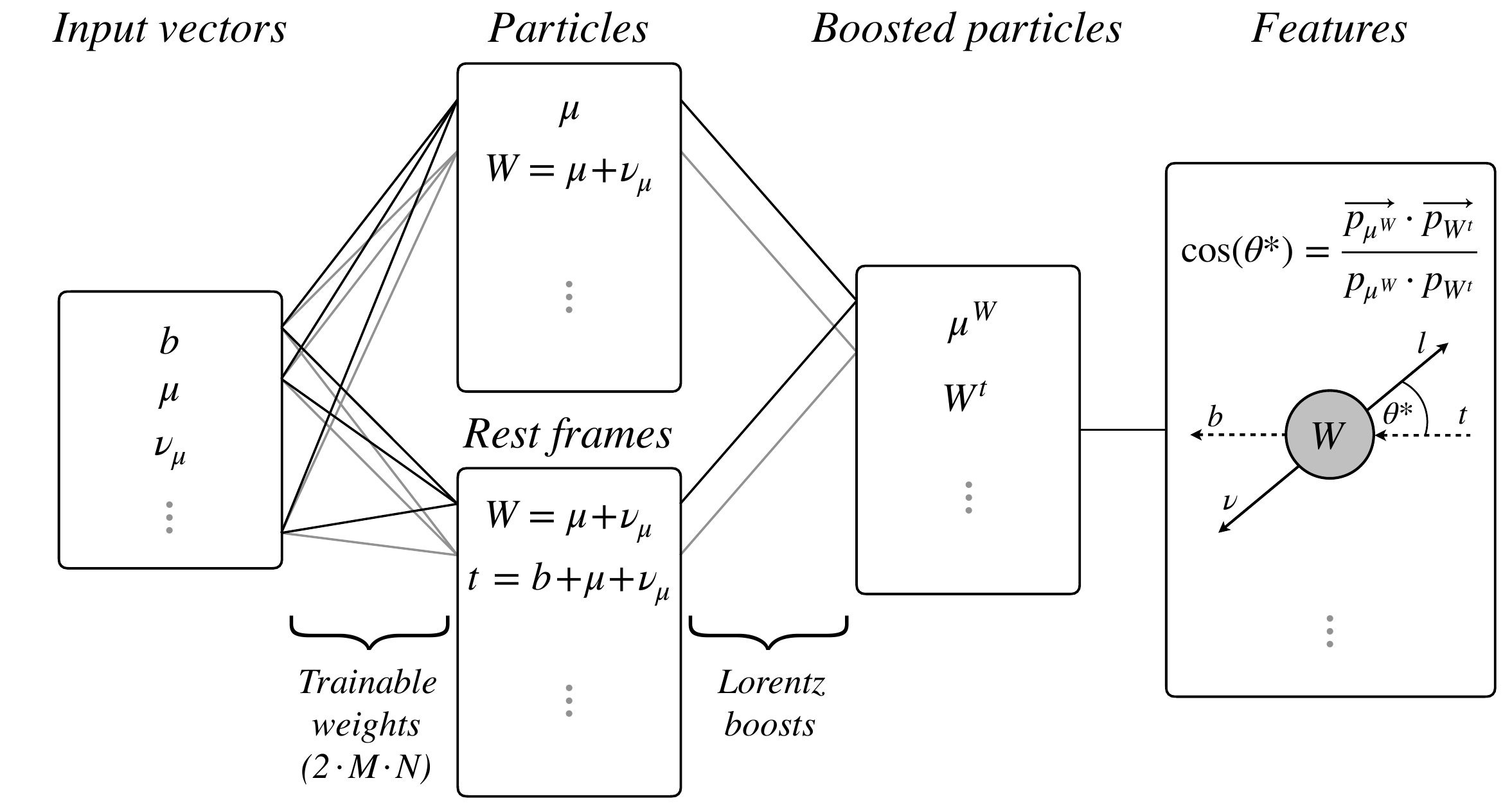}
    \caption{
        Example of a possible feature engineering in top quark decays addressing the angular distance of the direction of the W boson in the top rest system and the direction of the lepton in the W boson rest system, commonly referred to as $\cos(\theta^*)$.
    }
    \label{fig:lbn_example}
\end{figure}

Examples of variables that can be reconstructed with this structure include spin-dependent angular distributions, such as those observed during the decay of a top quark with subsequent leptonic decay of the W boson.
By boosting the charged lepton into the rest frame of the W boson, its decay angular distribution can be investigated.
In more sophisticated scenarios, the LBN is also capable of accessing further properties that rely on the characteristics of two different rest frames.
An example is a variable usually referred to as $\cos(\theta^*)$, which is defined by the angular difference between the directions of the charged lepton in the W boson's rest frame, and the W boson in the rest frame of the top quark.
The procedure of how this variable can be reconstructed in the LBN is depicted in \Fig{fig:lbn_example}.

The number $N$ of incoming particles is to be chosen according to the research question.
The number of matching combinations $M$ is a hyperparameter to be adjusted.
In this paper we introduce a specific version of the LBN which constructs $M$ particle combinations, and each combination will have its own suitable rest system.
Other variants are conceivable and will be mentioned below.

In the following paragraphs we describe the network architecture in detail.

\subsection*{Combinations}
\label{sec:architecture:combinations}

The purpose of the first LBN layer is to construct arbitrary particles and suitable rest frames for subsequent boosting.
The construction is realized via linear combinations of $N$ input four-vectors,
\begin{equation}
    X =
    \begin{bmatrix}
        E_1    & p_{x,1} & p_{y,1} & p_{z,1}\\
        E_2    & p_{x,2} & p_{y,2} & p_{z,2}\\
        \vdots & \vdots  & \vdots  & \vdots\\
        E_N    & p_{x,N} & p_{y,N} & p_{z,N}
    \end{bmatrix},
\end{equation}
to a number of $M$ particles and rest frames, which are represented by four-vectors accordingly.
Here, $M$ is a hyperparameter of the LBN and its choice is related to the respective physics application.
The coefficients $W$ of all linear combinations $C$,
\begin{equation}
    C_{m} = \sum_{n=1}^N W_{mn} \cdot X_{n}
\end{equation}
with $m \in \left[1, M\right]$, are free parameters and subject to optimization within the scope of the training process.
In the following, combined particles and rest frames are referred to as $C^P$ and $C^R$, respectively.
Taking both into consideration, this amounts to a total of $2 \cdot N \cdot M$ degrees of freedom in the LBN.
In order to prevent the construction of objects which would lead to unphysical implications when applying Lorentz transformations, i.e., four-vectors not fulfilling $E > m > 0$, all parameters $W_{mn}$ are restricted to positive values.
We initialize the weights randomly according to a half-normal distribution with mean $0$ and standard deviation $1 / M$.
It should be noted that, in order to maintain essential physical properties and relations between input four-vectors, feature normalization is not applied at this point.

\subsection*{Lorentz transformation}
\label{sec:architecture:boost}

The boosting layer performs a Lorentz transformation of the combined particles into their associated rest frames.
The generic transformation of a four-vector $q$ is defined as $q^* = \Lambda \cdot q$ with the boost matrix
\begin{align}
    \Lambda &=
    \begin{bmatrix}
        \gamma           & -\gamma\beta n_x       & -\gamma\beta n_y       & -\gamma\beta n_z\\
        -\gamma\beta n_x & 1 + (\gamma - 1) n_x^2 & (\gamma - 1) n_x n_y   & (\gamma - 1) n_x n_z\\
        -\gamma\beta n_y & (\gamma - 1) n_y n_x   & 1 + (\gamma - 1) n_y^2 & (\gamma - 1) n_y n_z\\
        -\gamma\beta n_z & (\gamma - 1) n_z n_x   & (\gamma - 1) n_z n_y   & 1 + (\gamma - 1) n_z^2
    \end{bmatrix}.
    \label{eqn:lambda}
\end{align}
The relativistic parameters $\gamma = E / m$, $\vec{\beta} = \vec{p} / E$, and $\vec{n} = \vec{\beta} / \beta$ are to be derived per rest-frame four-vector $C^R_m$.

The technical implementation within deep learning algorithms requires a vectorized representation of the Lorentz transformation.
To this end, we rewrite the boost matrix in \Eqn{eqn:lambda} as
\begin{align}
    \Lambda &= I \, + \, (U \oplus \gamma) \, \odot \, ((U \oplus 1) \cdot \beta \, - \, U) \, \odot \, (e \cdot e^T)\\[2mm]
    \text{with} \quad U &=
    \begin{bmatrix}
        -1^{1 \times 1} & 0^{1 \times 3}\\
        0^{3 \times 1}  & -1^{3 \times 3}
    \end{bmatrix},
    \quad e =
    \begin{bmatrix}
        1^{1 \times 1}\\
        -\vec{n}^{3 \times 1}
    \end{bmatrix},
\end{align}
and the $4 \times 4$ unit matrix $I$.
The operators $\oplus$ and $\odot$ denote elementwise addition and multiplication, respectively.
This notation also allows for the extension by additional dimensions to account for the number of combinations $M$ and an arbitrary batch size.
As a result, all boosted four-vectors $B$ are efficiently computed by a single, broadcasted matrix multiplication, $B = \Lambda^R \cdot C^P$.

While the description above focuses on a pairwise mapping approach, i.e., particle $C^P_m$ is boosted into rest frame $C^R_m$, other configurations are conceivable:
\begin{itemize}
    \item
    The input four-vectors are combined to build $M$ particles and $K$ rest frames.
    Each combined particle is transformed into all rest frames, resulting in $K \cdot M$ boosted four-vectors.

    \item
    The input four-vectors are combined only to build $M$ particles, which simultaneously serve as rest frames.
    Each combined particle is transformed into all rest frames derived from the other particles, resulting in $M^2 - M$ boosted four-vectors.
\end{itemize}
The specific advantages of these configurations can, in general, depend on aspects of the respective physics application.
Results of these variants will be the target of future investigations.

\subsection*{Feature extraction}
\label{sec:architecture:features}

Following the boosting layer, features are extracted from the Lorentz transformed four-vectors, which can then be utilized in a subsequent neural network.
For the projection of $M \times 4$ particle properties into $F \times 1$ features, we employ a distinct yet customizable set of generic mappings.
The autonomy of the network is not about finding entirely new features, but rather about factorizing probability densities in the most effective way possible to answer a scientific question.
Therefore, we let the LBN network work autonomously to find suitable particle combinations and rest frames which enable this factorization, but then use established particle characterizations.

We differentiate between two types of generic mappings:
\begin{enumerate}
    \item
    Single features are extracted per boosted four-vector.
    Besides the vector elements ($E$, $p_x$, $p_y$, $p_z$) themselves, derived features such as mass, transverse and longitudinal momentum, pseudorapidity, and azimuth can be derived.

    \item
    Pairwise features are extracted for all pairs of boosted four-vectors.
    Examples are the cosine of their spatial angular difference, their distance in the $\eta-\phi$ plane, or their distance in Minkowski space.
    In contrast to single features, pairwise features introduce close connections among boosted four-vectors and, by means of backpropagation, between trainable combinations of particles and rest frames.
\end{enumerate}
The set of extracted features is concatenated to a single output vector.
Provided that the employed batch size is sufficiently large, batch normalization with floating averages adjusted during training can be performed after this layer \cite{batch_norm}.

\subsection*{Subsequent problem-specific application}
\label{sec:architecture:application}

In the previous sections, we described the LBN as shown in the left box in \Fig{fig:lbn_arch}, namely how input vectors are combined and boosted into dedicated rest frames, followed by how features of these transformed four-vectors are compiled.
These features are intended to maximize the information content to be exploited in a subsequent, problem-specific deep neural network NN as shown in the right box in \Fig{fig:lbn_arch}.

The objective function of the NN defines the type of learning process.
Weight updates are performed as usual through backpropagation.
These updates apply to the weights of the subsequent network as well as to the trainable weights of the combination layer in the LBN.

In the following, we evaluate how well the autonomous feature engineering of the compound LBN+NN network operates.
We compare its performance with only low-level information to that of typical, fully-connected deep neural networks (DNNs).
We alternatively supply the DNNs with low-level information, sophisticated high-level variables (see \Apx{sec:appendix}), and the combination of both.

\section{Simulated datasets}
\label{sec:simulations}

\begin{figure}[h!tbp]
    \centering
    \begin{subfigure}{0.5\textwidth}
        \centering
        \includegraphics[width=0.7\textwidth]{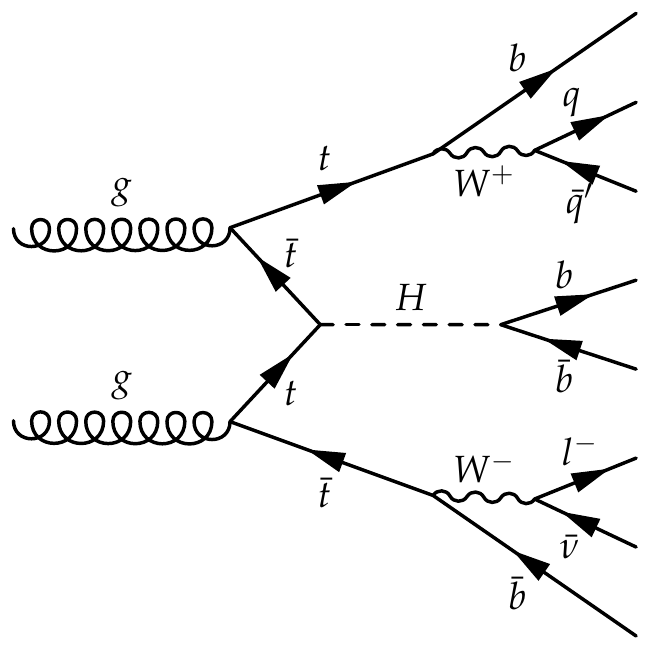}
        \caption{}
        \label{fig:feynman_ttH}
    \end{subfigure}%
    \begin{subfigure}{.5\textwidth}
        \centering
        \includegraphics[width=0.7\textwidth]{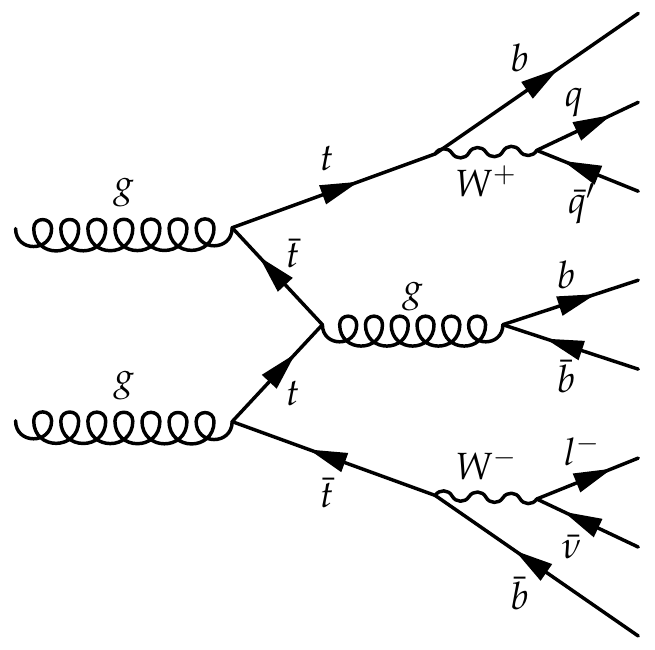}
        \caption{}
        \label{fig:feynman_ttbb}
    \end{subfigure}
    \caption{Example Feynman diagrams of a) $\ttH$ and b) $\ttbb$ processes.}
    \label{fig:feynman}
\end{figure}

The Pythia $8.2.26$ program package \cite{Pythia} was used to simulate $\ttH$ and $\ttbb$ events.
Examples of corresponding Feynman diagrams are shown in \Fig{fig:feynman}.
The matrix elements contain angular correlations of the decay products from heavy resonances.
Beam conditions correspond to LHC proton-proton collisions at $\sqrt{s} = 13$\,TeV.
Only the dominant gluon-gluon process is enabled and Higgs boson decays into bottom quark pairs are favored.
Hadronization is performed with the Lund string fragmentation model.

To analyze a typical final state as observed in an LHC detector, we use the DELPHES package \cite{deFavereau:2013fsa}.
DELPHES provides a modular framework designed to parameterize the simulation of a multi-purpose detector.
In our study, we chose to simulate the CMS detector response.
All important effects such as pile-up, deflection of charged particles in magnetic fields, electromagnetic and hadronic calorimetry, and muon detection systems are covered.
The simulated output consists of muons, electrons, tau leptons, photons, jets, and missing transverse momentum from a particle flow algorithm.
As the neutrino leaves the detector without interacting, we identify its transverse momentum components from the measurement of missing transverse energy, and we reconstruct its longitudinal momentum by constraining the mass of the leptonically decaying W boson to $80.4$\,GeV.

Following the detector simulation, an event selection with the following criteria was carried out:
Events must have at least six jets clustered with the anti-$k_T$ algorithm, implemented in the FastJet package \cite{Cacciari:2011ma}, with a radius parameter of $\Delta R = 0.4$.
A jet is accepted if its transverse momentum fulfills $p_t > 25$\,GeV and its absolute pseudorapidity is within $|\eta| < 2.4$.
Furthermore, exactly one electron or muon with $p_t > 20$\,GeV and $|\eta| < 2.1$ is required.
Events with further leptons are rejected to focus solely on semi-leptonic decays of the $\ttbar$ system.
Finally, a matching is performed between the final-state partons of the generator and the jets of the detector simulation with the maximum distance $\Delta R = 0.3$, whereby the matching must be unambiguously successful for all partons.
For $\ttH$ events, the combined selection efficiency was measured as $2.3$\,\%.

For $\ttbb$ processes, further measures are taken to identify the two additional bottom quark jets.
The definition is an adaption of \cite{Sirunyan:2018mvw}.
At generator level, the anti-$k_T$ algorithm with a radius parameter of $\Delta R = 0.4$ is used to cluster all stable final-state particles.
If a generator jet is found to originate from one of the top quarks or W boson decays, or if its transverse momentum is below a threshold of $20$\,GeV, it is excluded from further considerations.
For each remaining jet, we then count the number of contained bottom hadrons using the available hadronization and decay history.
An event is a $\ttbb$ candidate if at least two generator jets contain one or more distinct bottom hadrons.
Finally, a matching is performed between the four quarks resulting from the $\ttbar$ decay and the two identified generator jets on the one hand, and the selected, reconstructed jets on the other hand.
Similar to $\ttH$, a $\ttbb$ event is accepted if all six generator-level objects are unambiguously matched to a selected jet.
This leaves a fraction of $0.02$\,\% of the generated $\ttbb$ events.

A total of $10^6$ events remain, consisting of an evenly distributed number of $\ttH$ and $\ttbb$ events, which are then divided into training and validation datasets at a ratio of $80:20$.

For further analysis, a distinct naming scheme for reconstructed jets is introduced that is inspired by the semi-leptonic decay characteristics of the $\ttbar$ system.
The two light quark jets of the hadronically decaying W boson are named $\qi$ and $\qii$, where the index $1$ refers to the jet with the greater transverse momentum.
The bottom jet that is associated to this W boson within a top quark decay is referred to as $\bhad$.
Accordingly, the bottom quark jet related to the leptonically decaying W boson is referred to as $\blep$.
The remaining jets are named $\bi$ and $\bii$, and transparently refer to the decay products of the Higgs boson for $\ttH$, or to the additional b quark jets for $\ttbb$ events.

\begin{figure}[h!tbp]
    \centering
    \begin{subfigure}{0.5\textwidth}
        \centering
        \includegraphics[width=0.9\textwidth]{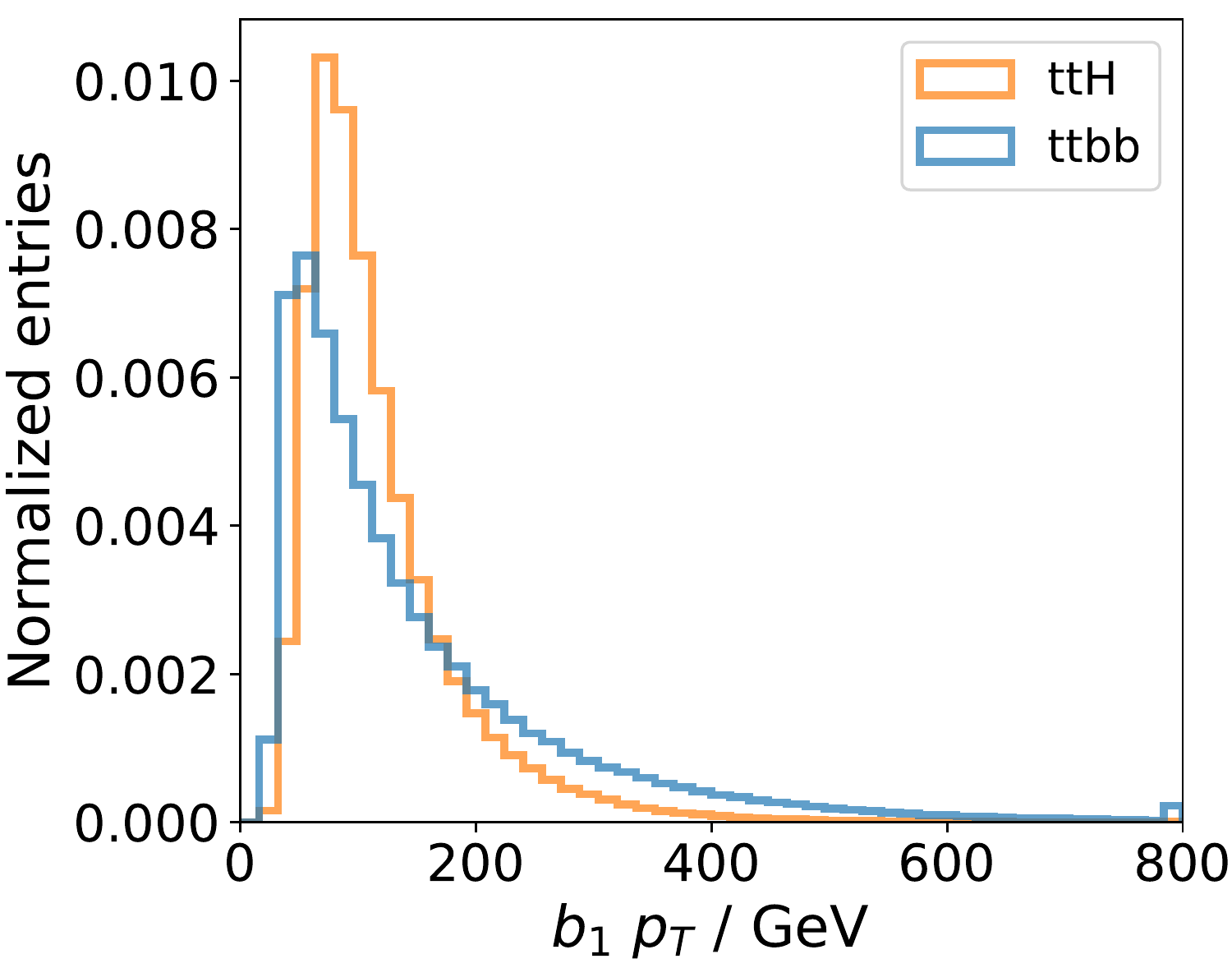}
        \caption{}
        \label{fig:dataset-a}
    \end{subfigure}%
    \begin{subfigure}{.5\textwidth}
        \centering
        \includegraphics[width=0.9\textwidth]{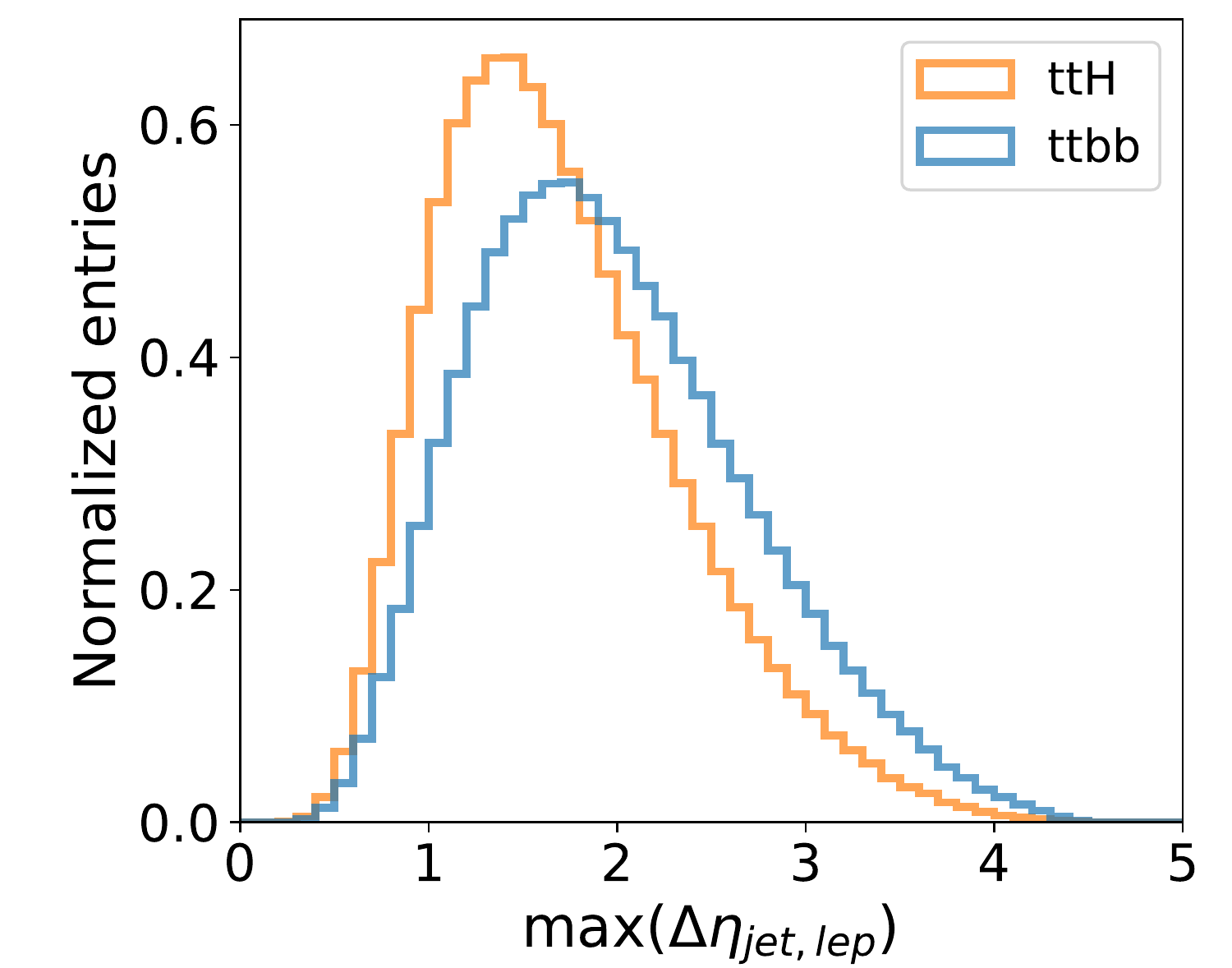}
        \caption{}
        \label{fig:dataset-b}
    \end{subfigure}
    \begin{subfigure}{0.5\textwidth}
        \centering
        \includegraphics[width=0.9\textwidth]{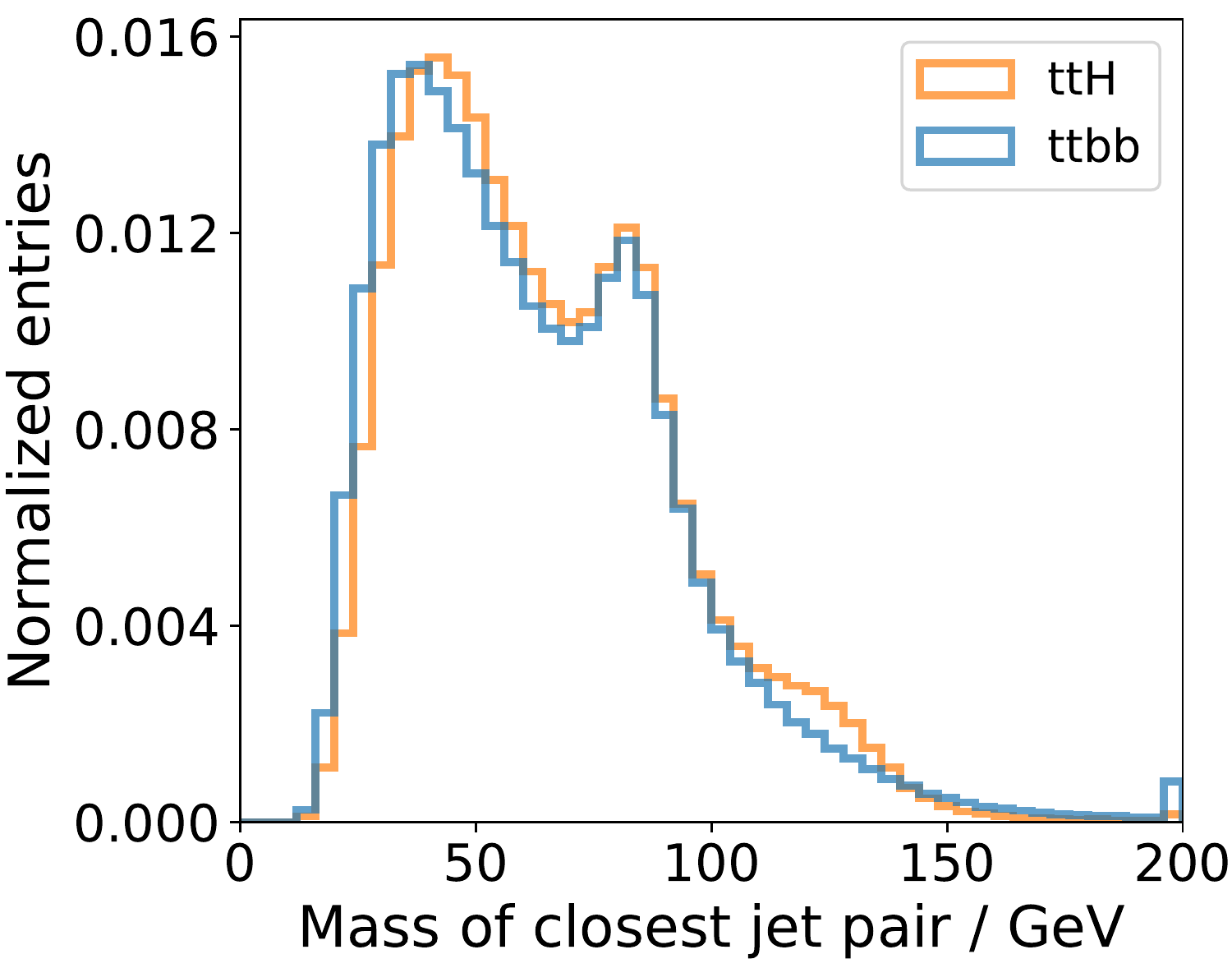}
        \caption{}
        \label{fig:dataset-c}
    \end{subfigure}%
    \begin{subfigure}{.5\textwidth}
        \centering
        \includegraphics[width=0.9\textwidth]{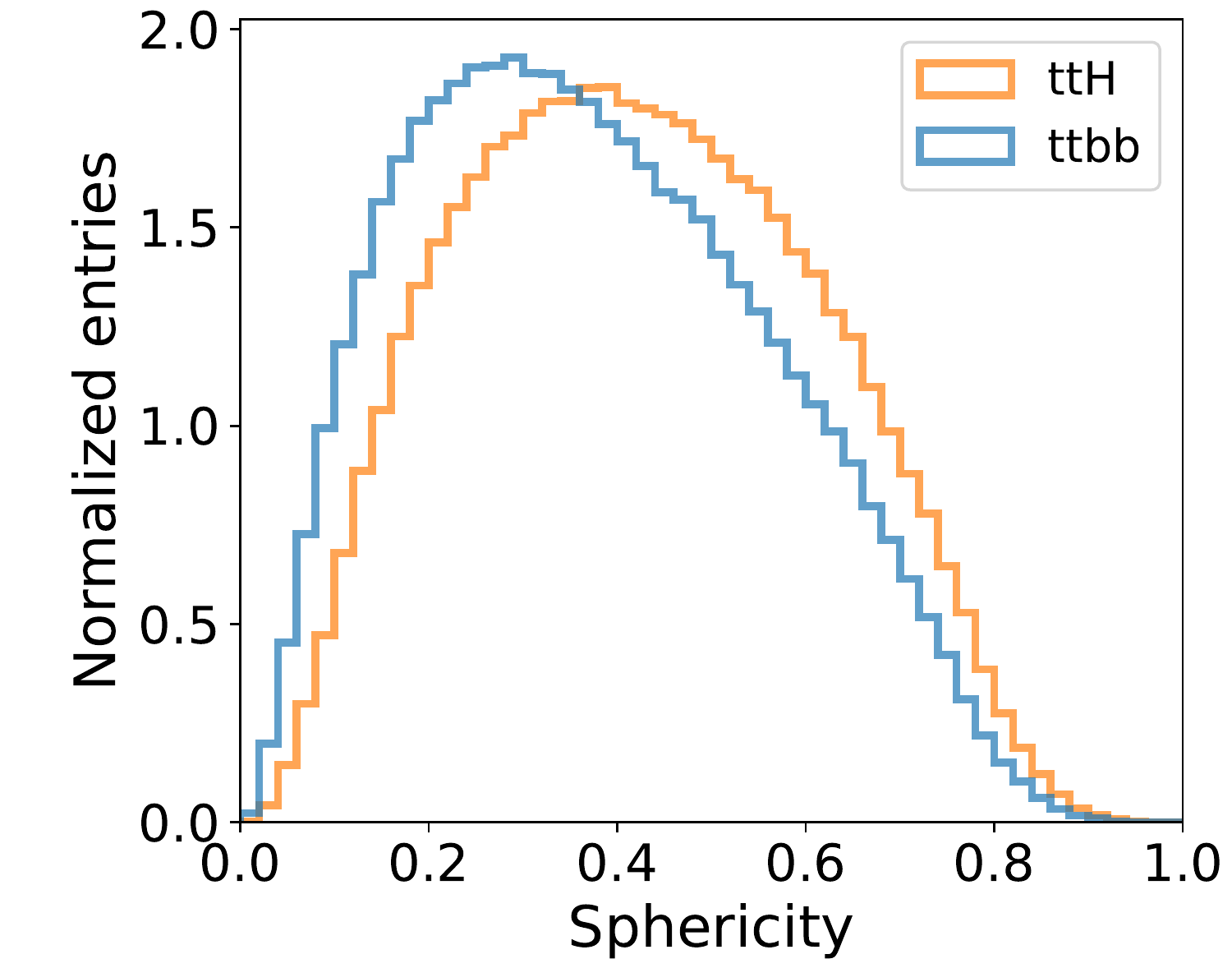}
        \caption{}
        \label{fig:dataset-d}
    \end{subfigure}
    \caption{
        Exemplary comparisons of kinematic distributions of $\ttH$ and $\ttbb$ events; a)~transverse momentum of the jet $\bi$, b)~largest difference in pseudorapidity of a jet to the charged lepton, c)~invariant mass of the closest jet pair, d)~event sphericity.
    }
    \label{fig:dataset}
\end{figure}

In \Fig{fig:dataset} we show, by way of example, distributions of the generated $\ttH$ and $\ttbb$ events.
In \Fig{fig:dataset-a} we compare the transverse momentum of the jet $\bi$, and in \Fig{fig:dataset-b} the largest difference in pseudorapidity between a jet and the charged lepton.
In \Fig{fig:dataset-c} we show the invariant mass of the closest jet pair, and in \Fig{fig:dataset-d} the event sphericity.

\section{Benchmarks of network performance}
\label{sec:performance}

As a benchmark, the LBN is utilized in the classification task of distinguishing a signal process from a background process.
In this example, we use the production of a top quark pair in association with a Higgs boson decaying into bottom quarks ($\ttH$) as a signal process, and top quark pairs produced with $2$ additional bottom jets ($\ttbb$) as a background process.
In both processes, the final state consists of eight four-vectors, four of which represent bottom quark jets, two describe light quark jets, one denotes a charged lepton, and one a neutrino.

Within the benchmark, the LBN competes against other frequently used deep neural network setups, which we refer to as DNN here.
For a meaningful comparison, we perform extensive searches for the optimal hyperparameters in each setup.
As explained above, the LBN consists of only very few parameters to be trained in addition to those of the following neural network (LBN+NN).
Still, by varying the number $M$ of particle combinations to be constructed from the eight input four-vectors, and by varying the number $F$ and types of generated features, we performed a total of $346$ training runs of the LBN+NN setup, and a similar amount for the competing DNN.
The best performing architectures are listed in \Apx{sec:appendix}.

The LBN network exclusively receives the four-vectors of the eight particles introduced above, which shall be denoted by 'LBN low' in the following.
For the networks marked with DNN we use three variants of inputs.
In the first variant, the DNN receives only the four-vectors of the eight particles ('DNN-low').
For the second variant, physics knowledge is applied in the form of $26$ sophisticated high-level variables which are inspired by \cite{Sirunyan:2018mvw} and incorporate comprehensive event information ('DNN-high').
A list of these variables, such as the event sphericity \cite{event_shape_variables} and Fox-Wolfram moments \cite{fox_wolfram_moments}, can be found in the \Apx{sec:appendix}.
In the third variant, we provide the DNNs with both low-level and high-level variables as input ('DNN-combined').

In the first set of our benchmark tests, the task of input particle identification is bypassed by using the generator information in order to exclusively measure the performance of the LBN.
Jets associated to certain quarks through matching are consistently placed at the same position of the $N$ input four-vectors.
We quantify how well the event classification of the networks performs with the integral of the receiver operating characteristic curve (ROC AUC).

\begin{figure}[h!tbp]
    \centering
    \begin{subfigure}{0.5\textwidth}
        \centering
        \includegraphics[width=0.9\textwidth]{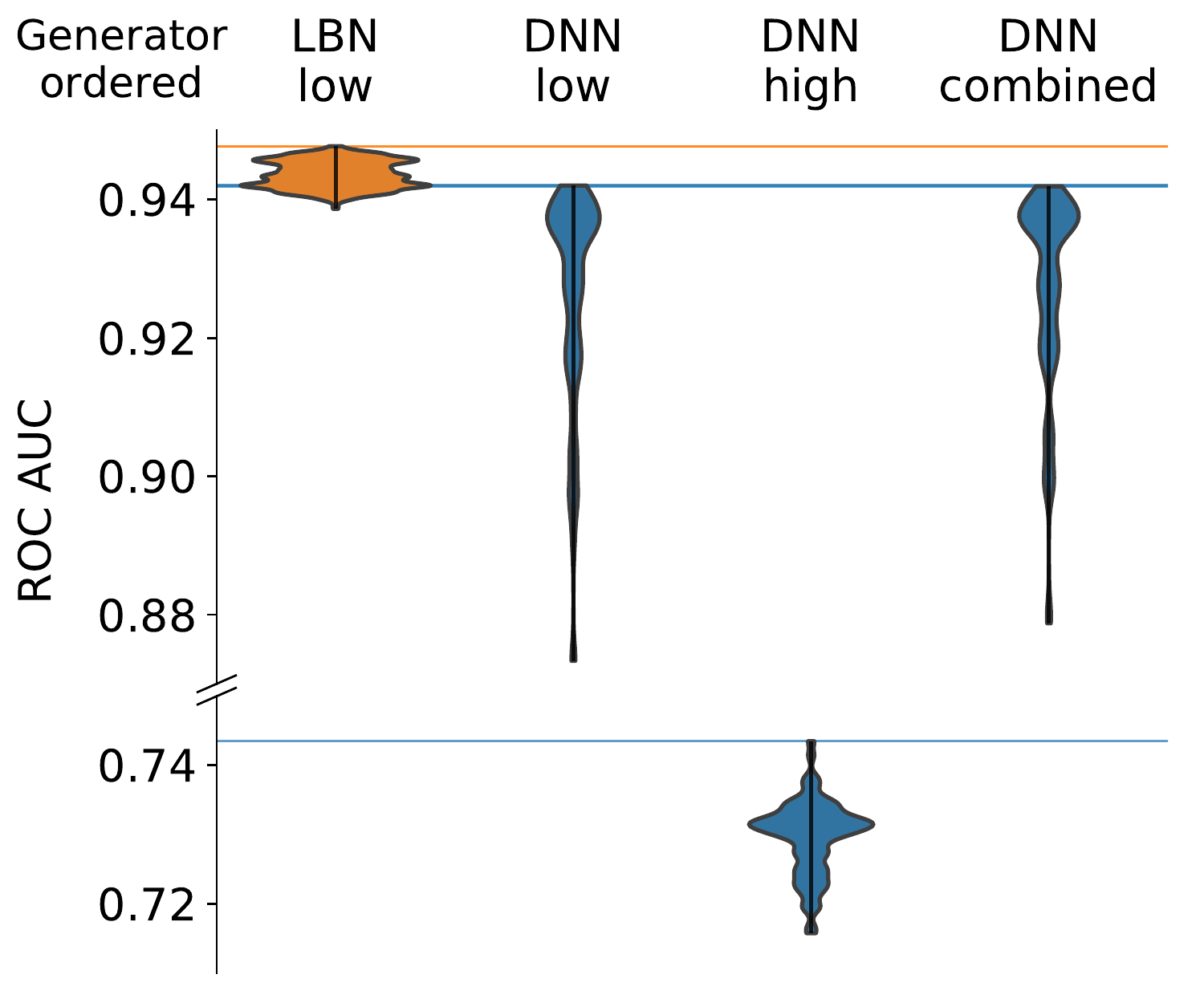}
        \caption{}
        \label{fig:perf_comparison_a}
    \end{subfigure}%
    \begin{subfigure}{0.5\textwidth}
        \centering
        \includegraphics[width=0.9\textwidth]{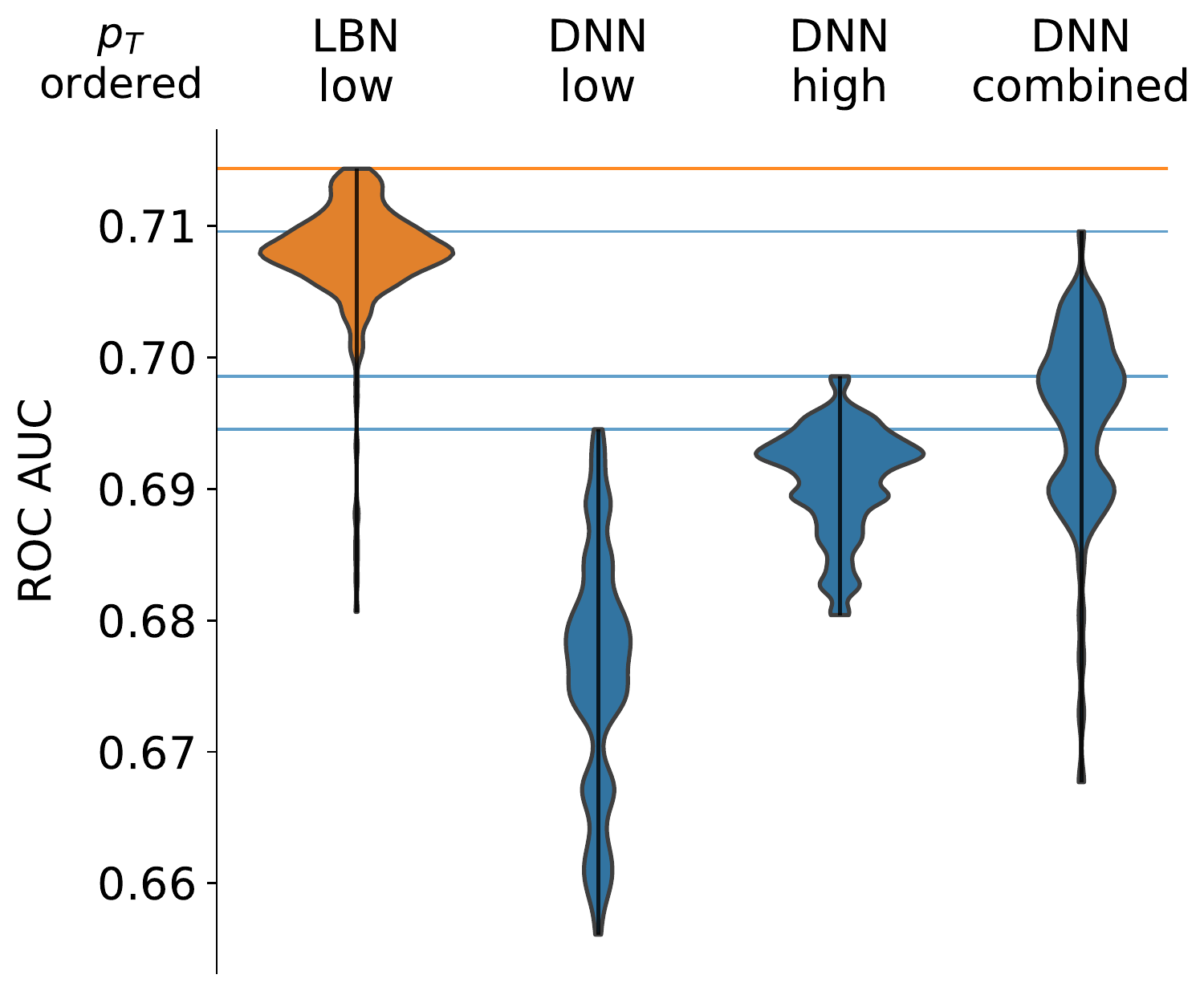}
        \caption{}
        \label{fig:perf_comparison_b}
    \end{subfigure}
    \caption{
        Benchmarks of the LBN performance in comparison to typical deep neural network architectures (DNNs) with three variants of input variables;
        low=four-vectors, high=physics-motivated variables, combined=low+high.
        Ordering of the input four-vectors is done by a) generator information and b) jet transverse momenta.
    }
    \label{fig:perf_comparison}
\end{figure}

\Fig{fig:perf_comparison_a} shows the performance of all training runs involved in the hyperparameter search.
The best training is marked with a horizontal line, while the distribution of results is denoted by the orange and blue shapes which manifestly depend on the structure of the scanned hyperparameter space.
For the LBN, the training runs achieve stable results since there are only minor variations in the resulting ROC AUC.

Also shown are the training runs of the three input variants for the DNNs.
If the eight input four-vectors are ordered according to the generator information, the 'DNN low' setup already achieves results equivalent to the combination of low-level and high-level variables ('DNN combined').
However, both setups are unable to match the 'LBN low' result.
Note that high-level variables are mostly independent of the input ordering by construction, and hence contain reduced information compared to the ordered four-vectors.
Consequently, the weaker performance of the 'DNN high' is well understood.

In the second set of our benchmark test, we disregard generator information and instead sort the six jets according to their transverse momenta, followed by the charged lepton and the neutrino.
This order can be easily established in measured data.
\Fig{fig:perf_comparison_b} shows that this alternative order consistently reduces the ROC AUC score of all networks.

The training runs of the LBN again show stable results with few exceptions and achieve the best performance in separating events from signal and background processes.
For the DNNs, the above-mentioned hierarchy emerges, i.e., the best results are achieved using the combination of high- and low-level variables, followed by only high-level variables with reasonably good results,
whereas the DNN exhibits the weakest performance with only low-level information.

Overall, the performance of the compound LBN+NN structure based on the four-vector inputs is quite convincing in this demanding benchmark test.

\section{Visualizations of network predictions}
\label{sec:insights}

Several components of the LBN architecture have a direct physics interpretation.
Their investigation can provide insights into what the network is learning.
In the following, we investigate only the best network trained on the generator-ordered input in order to establish which of the quark and lepton four-vectors are combined.

The number of particle combinations to create is a hyperparameter of the architecture.
For the generator ordering, we obtain the best results for $13$ particles.
This matches well with our intuition since the Feynman diagram also contains $13$ particles in the cascade.
In total we extract $F = 143$ features in the LBN that serve as input to the subsequent deep neural network NN (\Fig{fig:lbn_arch}).
For details refer to the \Apx{sec:appendix}.

In our particular setup, each combined particle has a separate corresponding rest frame.
As a consequence of demanding the second network to solve a research question, the decisions about which four-vectors are combined for the composite particle and which for its rest frame are strongly correlated.
A correlation also exists between all $13$ systems of combined particles and their rest frames from the pairwise angular distances exploited by the LBN as additional properties.

We start by looking at the weights used to combine the input four-vectors to form the particle combinations and rest frames.

\subsection*{Weight matrices of particle and rest-frame combinations}

\begin{figure}[h!tbp]
    \centering
    \begin{subfigure}{1\textwidth}
        \centering
        \includegraphics[width=0.65\textwidth]{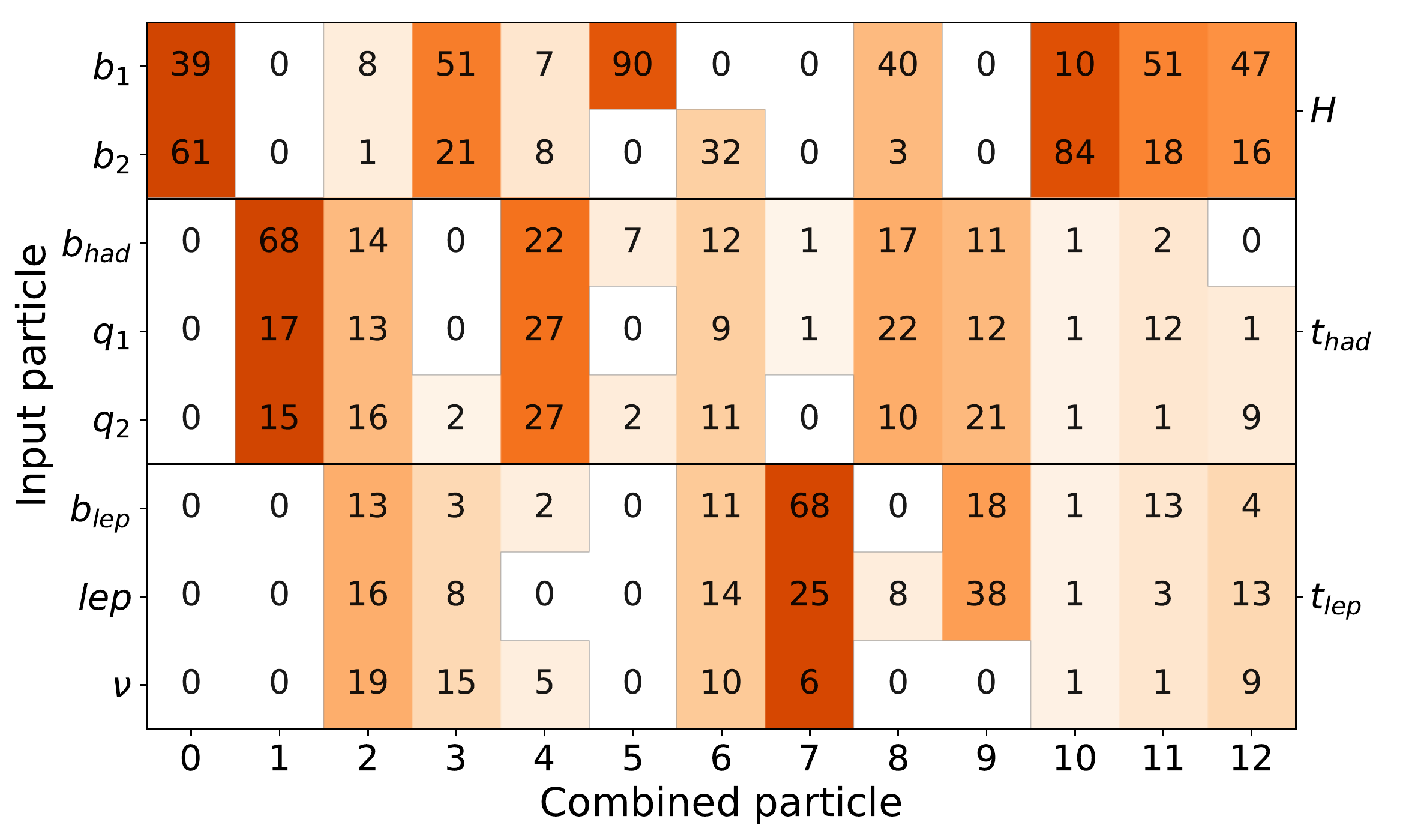}
        \caption{}
        \label{fig:weight_gen_particles_normed-a}
    \end{subfigure}
    \begin{subfigure}{1\textwidth}
        \centering
        \includegraphics[width=0.65\textwidth]{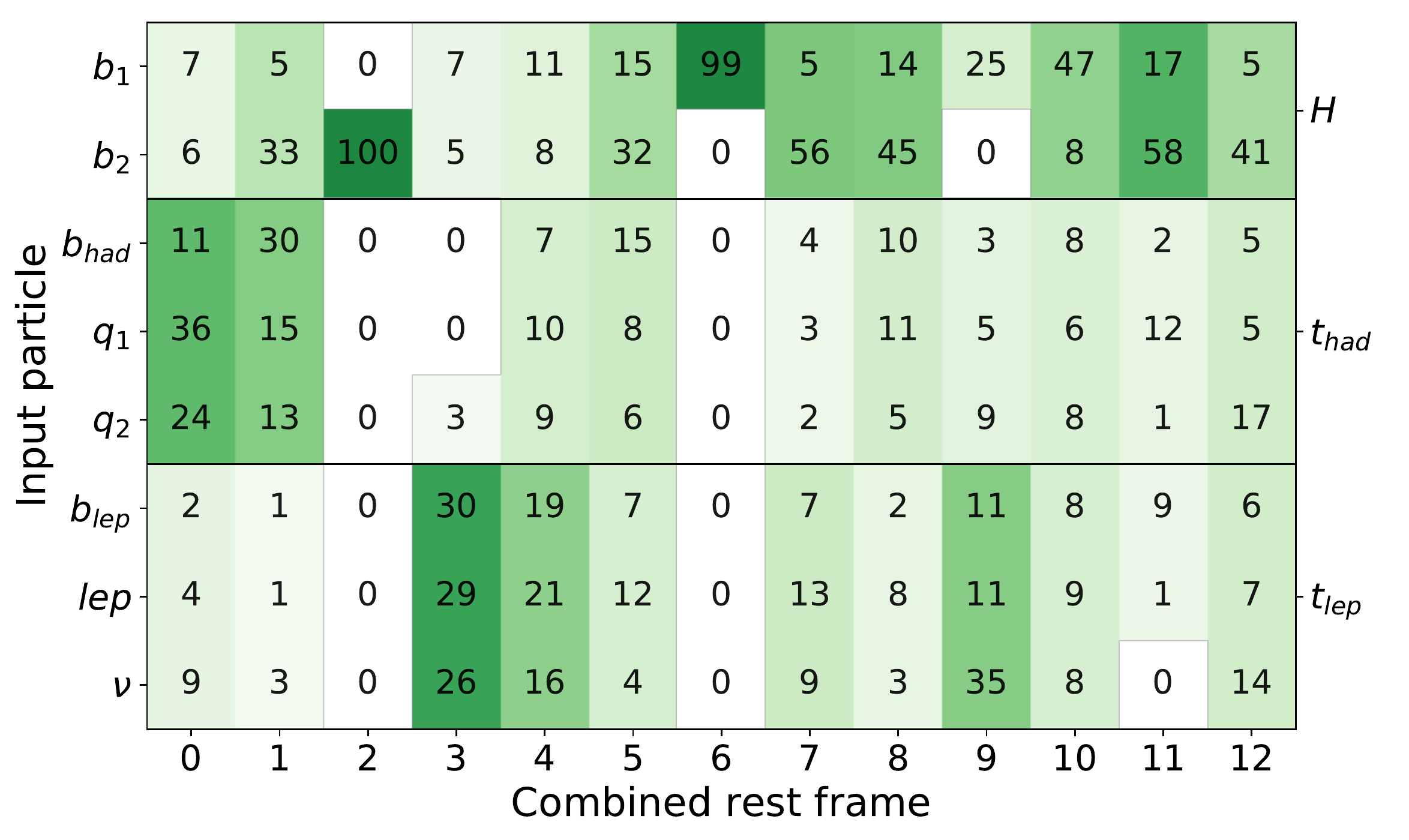}
        \caption{}
        \label{fig:weight_gen_particles_normed-b}
    \end{subfigure}
    \caption{
        Particle combinations (red) and their rest frames (green).
        The numbers represent the relative weights of the input particles for each combination $i$.
        The color code reflects the sum of the particle weights for the Higgs boson ($\ttH$), the $b\bar{b}$ system ($\ttbb$), or the two top quarks.
        For better clarity of the presentation, the zero weights are kept white.
    }
    \label{fig:weight_gen_particles_normed}
\end{figure}

\Fig{fig:weight_gen_particles_normed} shows the weight matrices of the LBN.
Each column of the two matrices for combined particles (red) and rest frames (green) is normalized so that the weight of each input four-vector (bottom quark jet $\bi$ of the Higgs, etc.) is shown as a percentage.
The color code reflects the sum of the particle weights for the Higgs boson (or $b\bar{b}$ system for $\ttbb$) and the two top quarks, respectively.

Note that combined particles and rest frames are complementary.
Extreme examples of this can be seen in columns $2$ and $6$, in which one of the two bottom quark jets from the Higgs boson decay was selected as the rest frame and a broad combination of all other four-vector vectors was formed whose properties are exploited for the event classification.

Conversely, in column $0$ a Higgs-like combination is transformed into the rest frame of all particles, to which the hadronic top quark makes the main contribution.
Similarly, in columns $1$ and $7$, top quark-like combinations are formed and also transformed into rest frames formed by many four-vectors.
These types of combinations allow for the Lorentz boost of the center-of-mass system of the scattering process to be nearly compensated.

\begin{figure}[htbp]
    \centering
    \begin{subfigure}{.49\textwidth}
        \centering
        \includegraphics[width=1\textwidth]{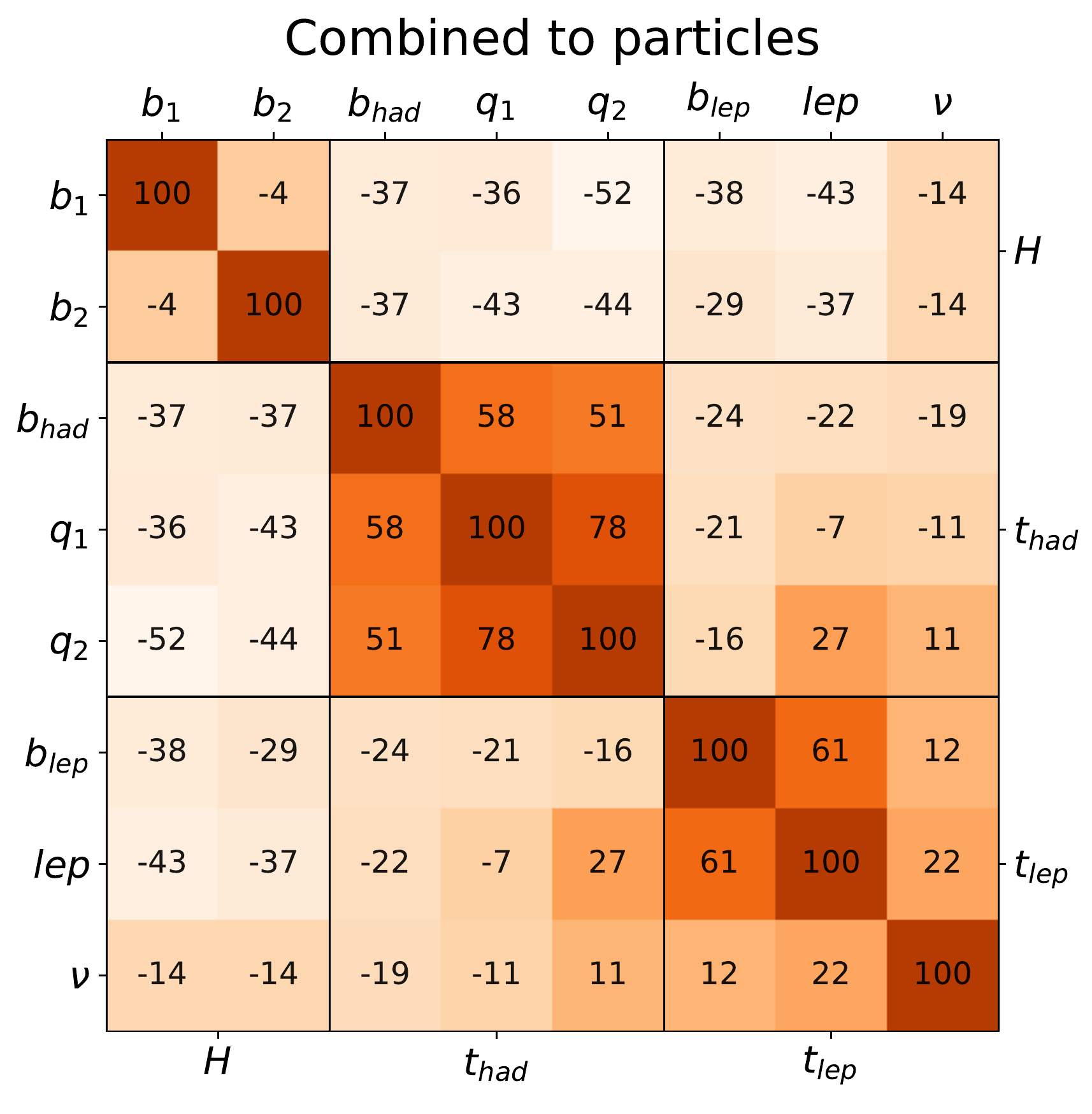}
        \caption{}
        \label{fig:correlations-a}
    \end{subfigure}
    \begin{subfigure}{.49\textwidth}
        \centering
        \includegraphics[width=1\textwidth]{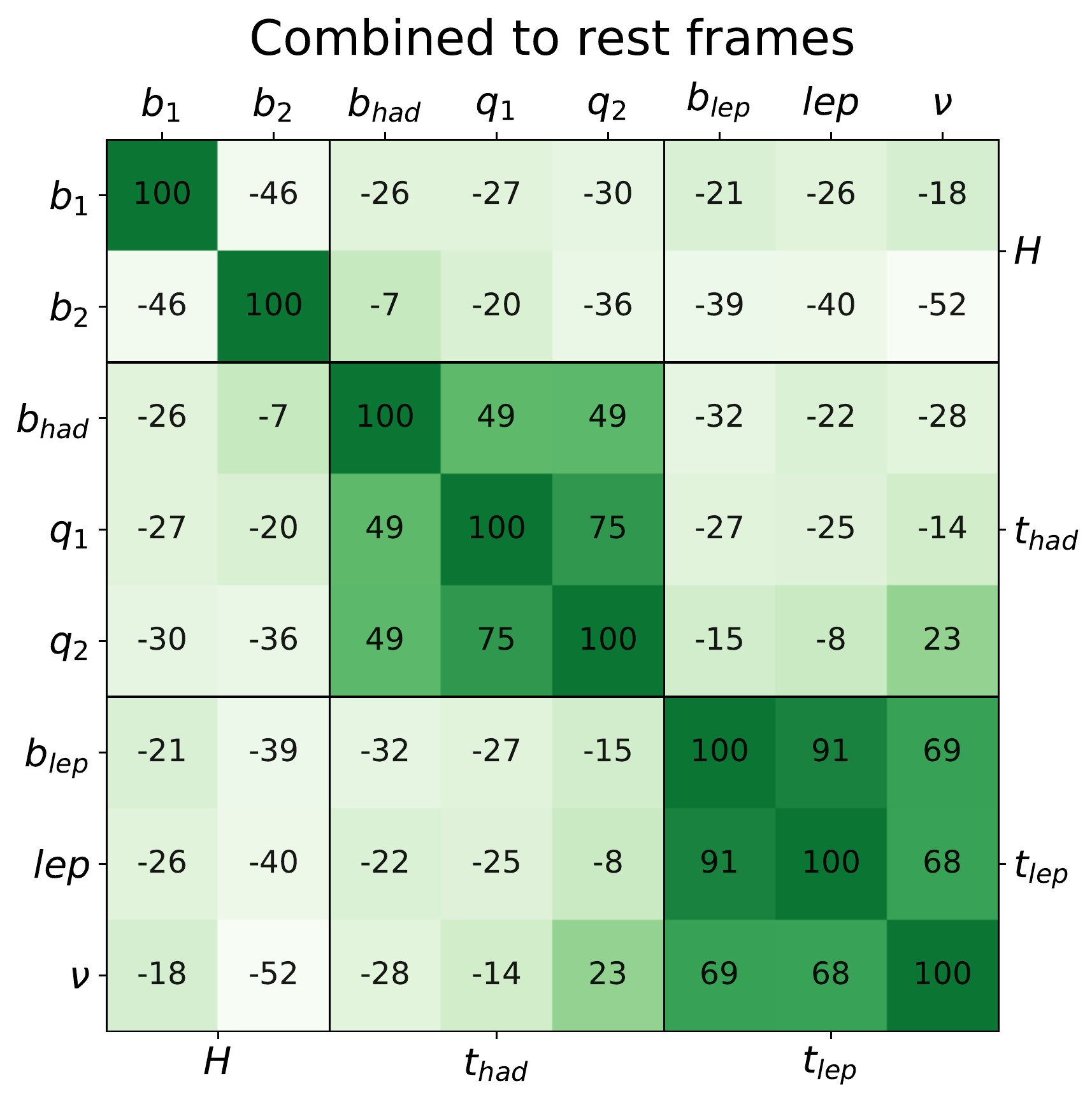}
        \caption{}
        \label{fig:correlations-b}
    \end{subfigure}
    \caption{
        Correlations of quarks and leptons for particle combinations (red) and for rest frames (green).
        Both the numbers and the color code represent the correlation strength.
    }
    \label{fig:correlations}
\end{figure}

It is striking that four-vector combinations forming a top quark are often selected.
In \Fig{fig:correlations}, we quantitatively assess the $13$ possible combinations of the combined particles (red) and the $13$ rest frames (green) in order to determine which combinations are typically formed between the eight input four-vectors.

In order to build rest frames (green), the LBN network recognizes combinations leading to the top quarks and treats the bottom quark jets from the Higgs individually.
This appears to be an appropriate choice as the top quark pair is the same in $\ttbb$ and $\ttH$ events while the two bottom quarks typically originate from gluon splitting or the Higgs decay, respectively.
For the hadronic top quark, the W boson ($75$\,\%) is often accounted for as well.
With the leptonic top, the LBN network considers bottom quark jet and lepton ($91$\,\%).
The lepton and neutrino are related to form the W ($68$\,\%).

To form the particle combinations (red), the LBN network builds the light quark jets to the W boson ($78$\,\%) and the hadronic top quark.
In the leptonic top quark, the LBN combines bottom quark jets and leptons ($61$\,\%).
Sometimes the lepton and the neutrino are also correlated to build the W boson ($22$\,\%).
The LBN rarely combines the Higgs boson ($-4$\,\%).

The positive and negative correlations show that the LBN forms physics-motivated particle combinations from the training data.
In the next step, we investigate the usefulness of these combinations for the separation of $\ttH$ and $\ttbb$ events.

\subsection*{Distributions resulting from combined particle properties}

The performance of particle combinations and their transformations into reference frames in the LBN is evaluated through the extracted features which support the classification of signal and background events.
Here, we present the distributions of different variables calculated from the combined particles.

\begin{figure}[h!tbp]
    \centering
    \begin{subfigure}{0.5\textwidth}
        \centering
        \includegraphics[width=0.9\textwidth]{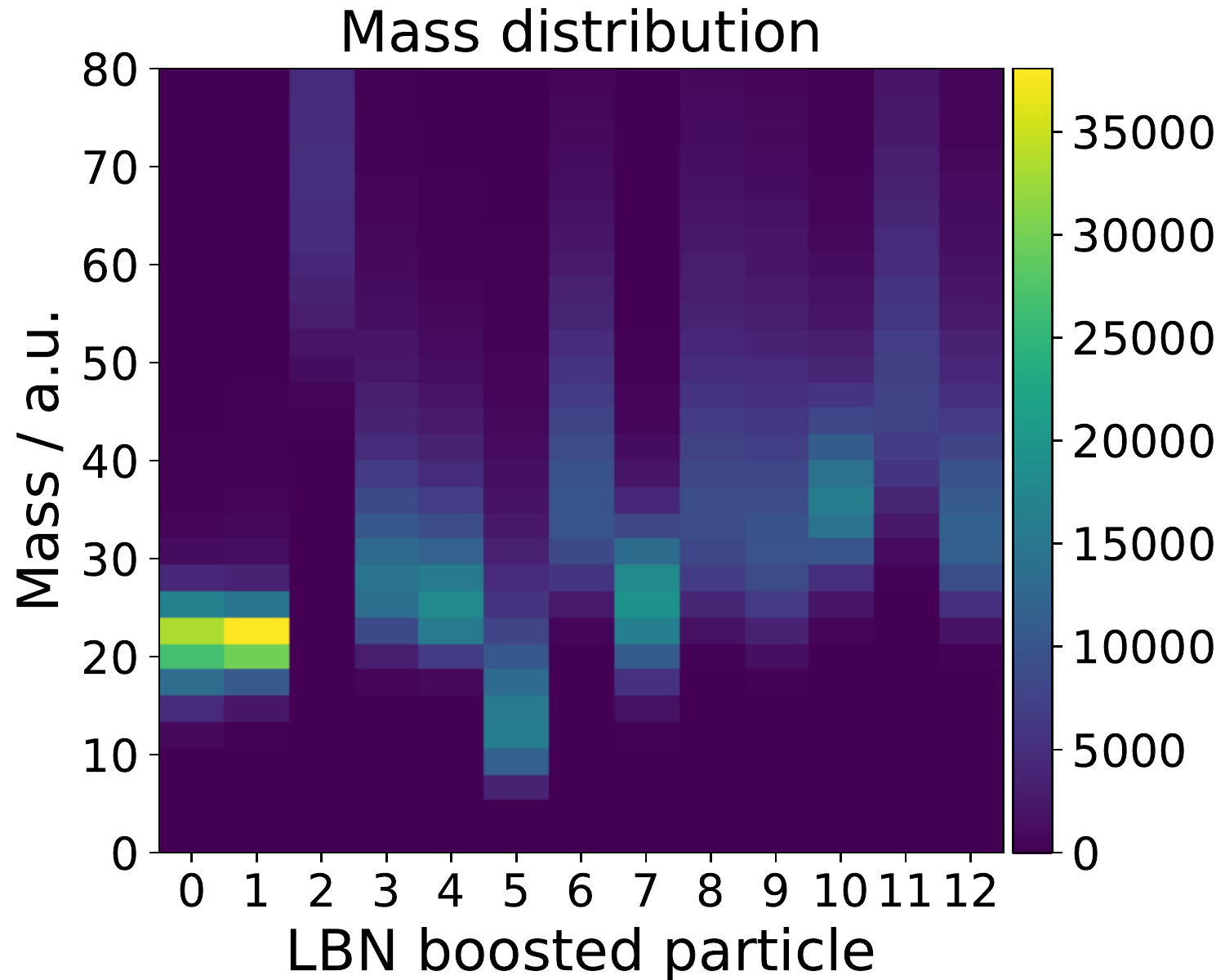}
        \caption{}
        \label{fig:m_gen_ttH_a}
    \end{subfigure}%
    \begin{subfigure}{.5\textwidth}
        \centering
        \includegraphics[width=0.9\textwidth]{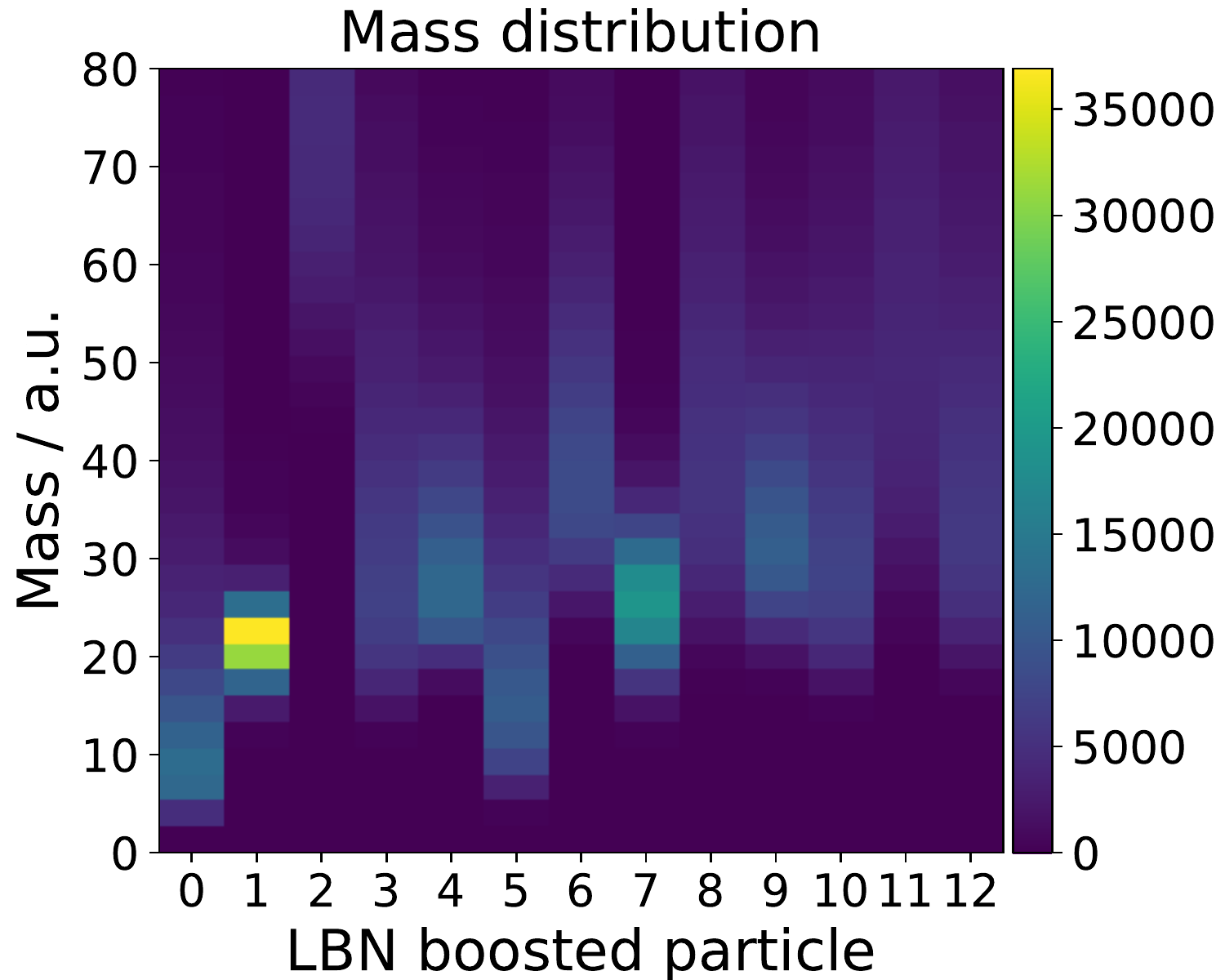}
        \caption{}
        \label{fig:m_gen_ttH_b}
    \end{subfigure}
    \begin{subfigure}{0.5\textwidth}
        \centering
        \includegraphics[width=0.9\textwidth]{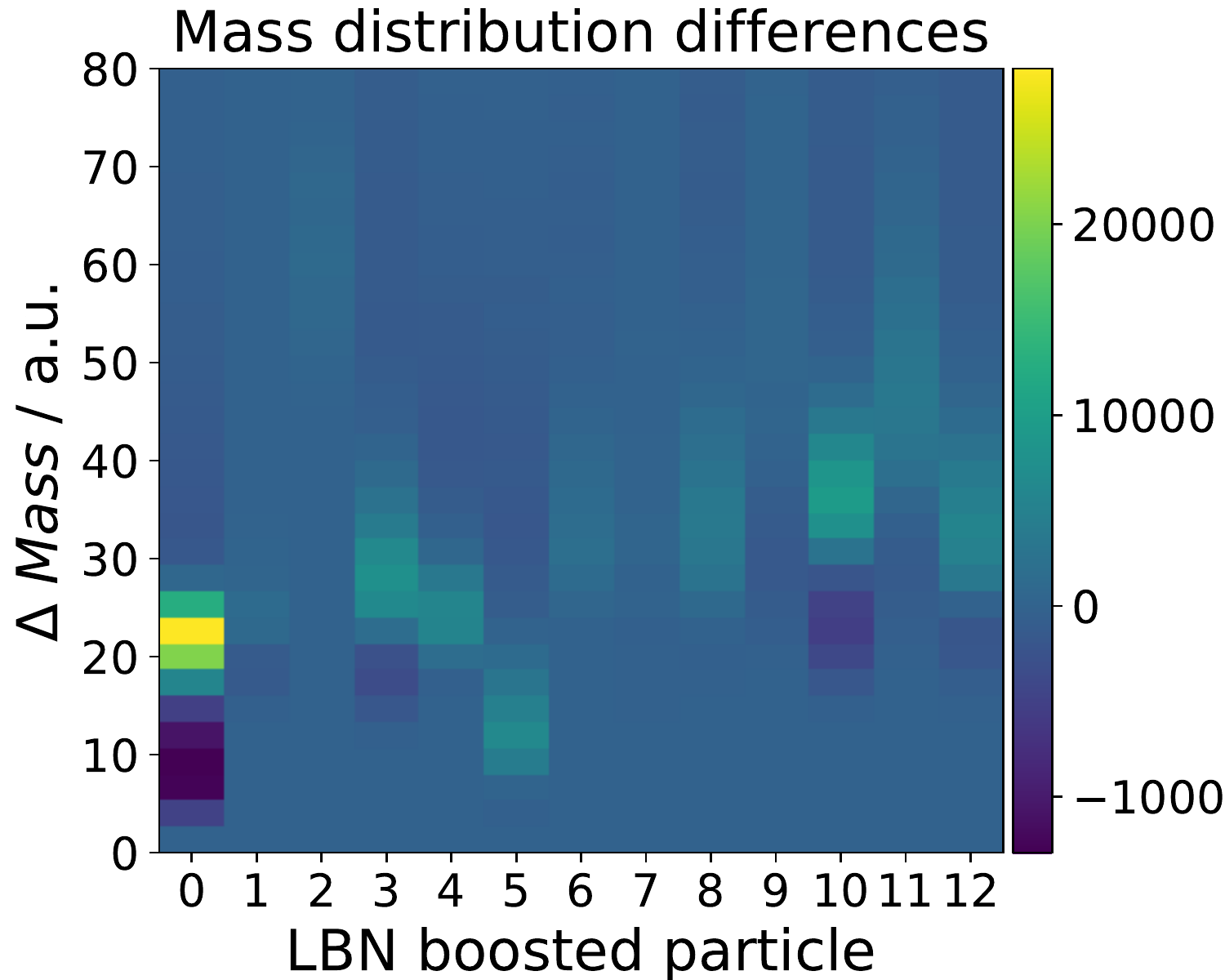}
        \caption{}
        \label{fig:m_gen_ttH_c}
    \end{subfigure}%
    \caption{
        Masses of combined particles for a) $\ttH$, b) $\ttbb$, and c) the differences between $\ttH$ and $\ttbb$.
    }
    \label{fig:m_gen_ttH}
\end{figure}

\Fig{fig:m_gen_ttH_a} shows the invariant mass $m$ distributions of all $13$ combined particles for the $\ttH$ signal dataset.
\Fig{fig:m_gen_ttH_b} shows the mass distributions for the $\ttbb$ background dataset.
The difference between the two histograms is shown in Figure~\ref{fig:m_gen_ttH_c}, where combined particle $0$ of the first column shows the largest difference between signal and background.
\Fig{fig:weight_gen_particles_normed} shows that this combination $0$ approximates a Higgs-like configuration (red).

\begin{figure}[h!tbp]
    \centering
    \begin{subfigure}{0.5\textwidth}
        \centering
        \includegraphics[width=0.8\textwidth]{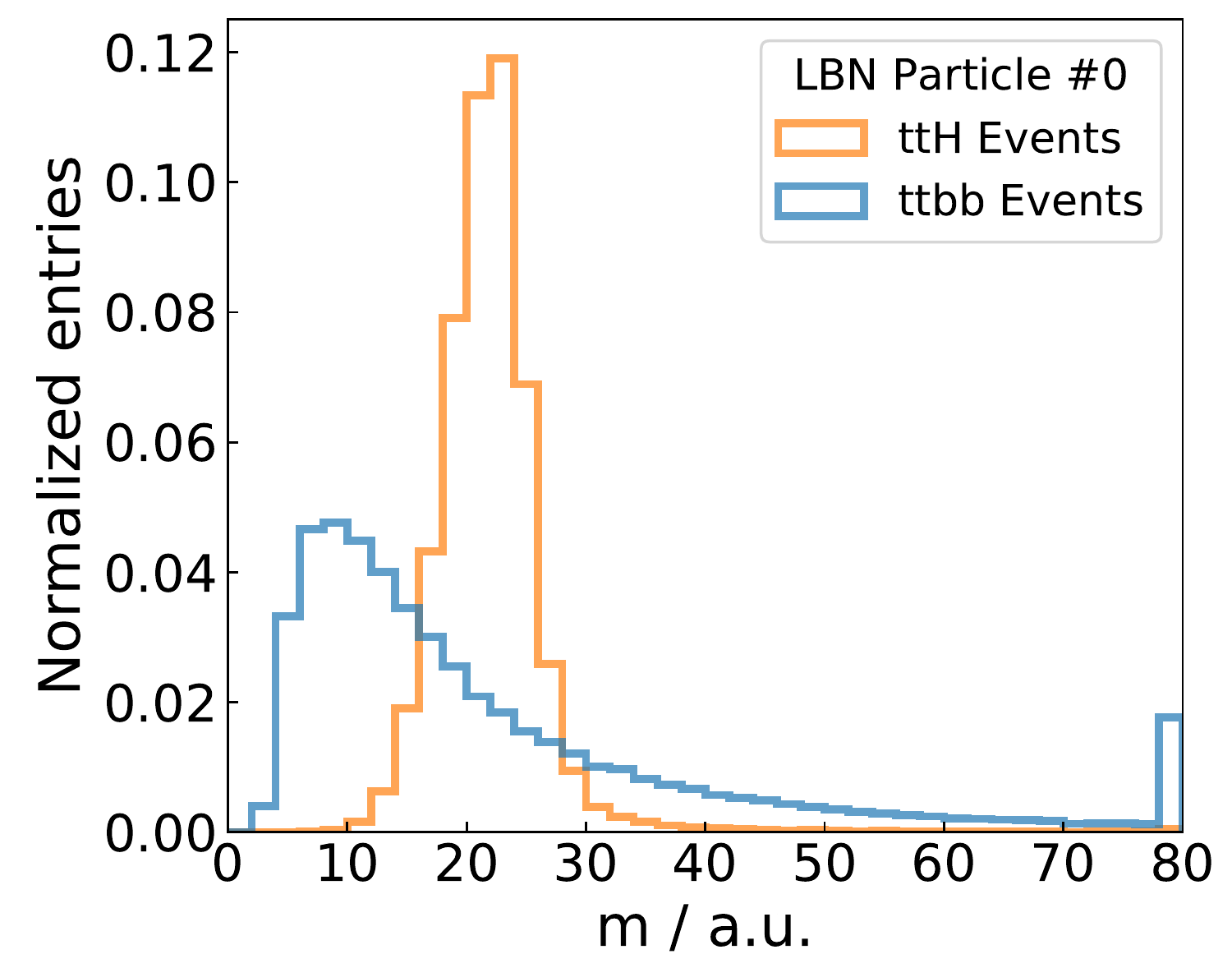}
        \caption{}
        \label{fig:slices_a}
    \end{subfigure}%
    \begin{subfigure}{.5\textwidth}
        \centering
        \includegraphics[width=0.8\textwidth]{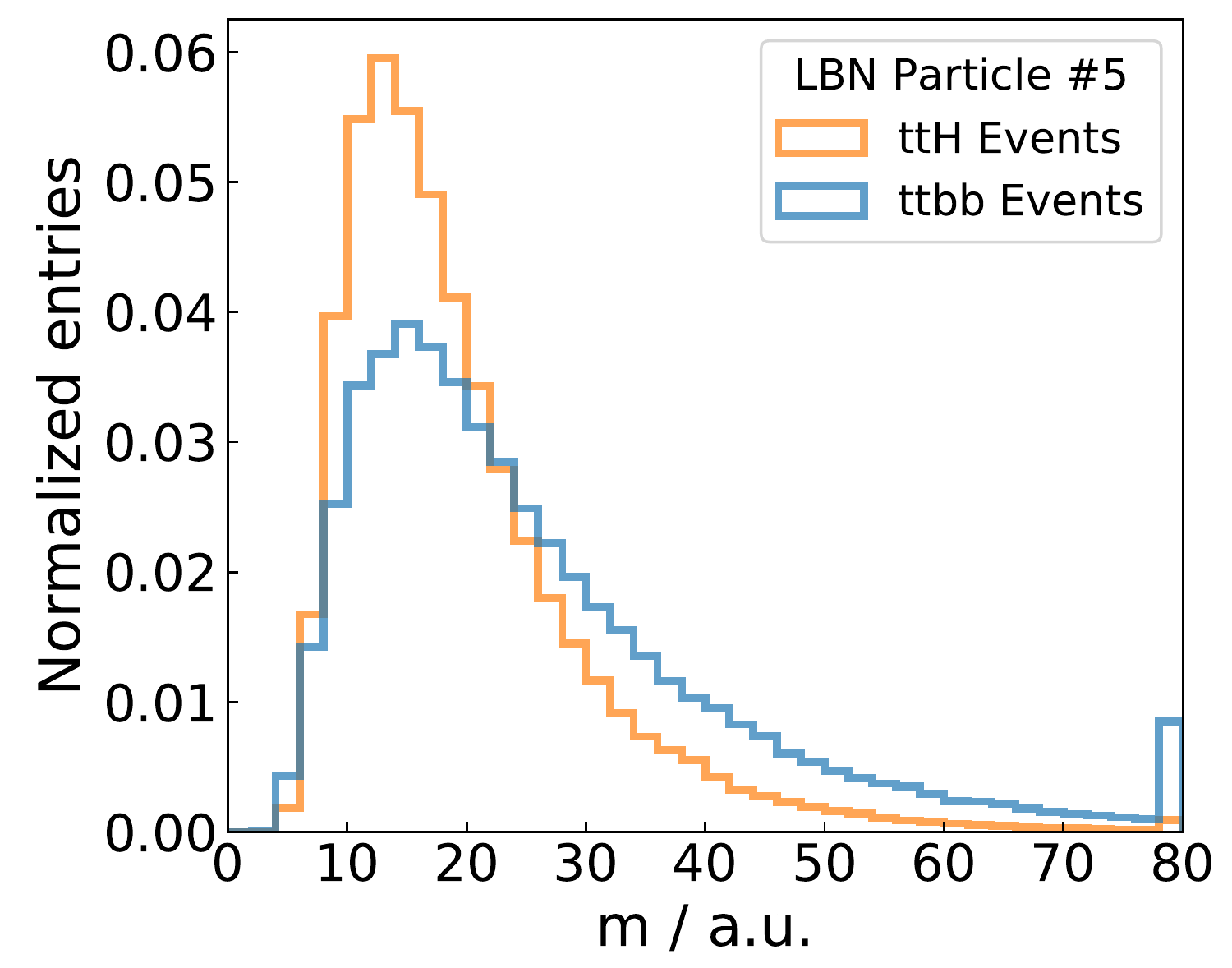}
        \caption{}
        \label{fig:slices_b}
    \end{subfigure}
    \begin{subfigure}{0.5\textwidth}
        \centering
        \includegraphics[width=0.8\textwidth]{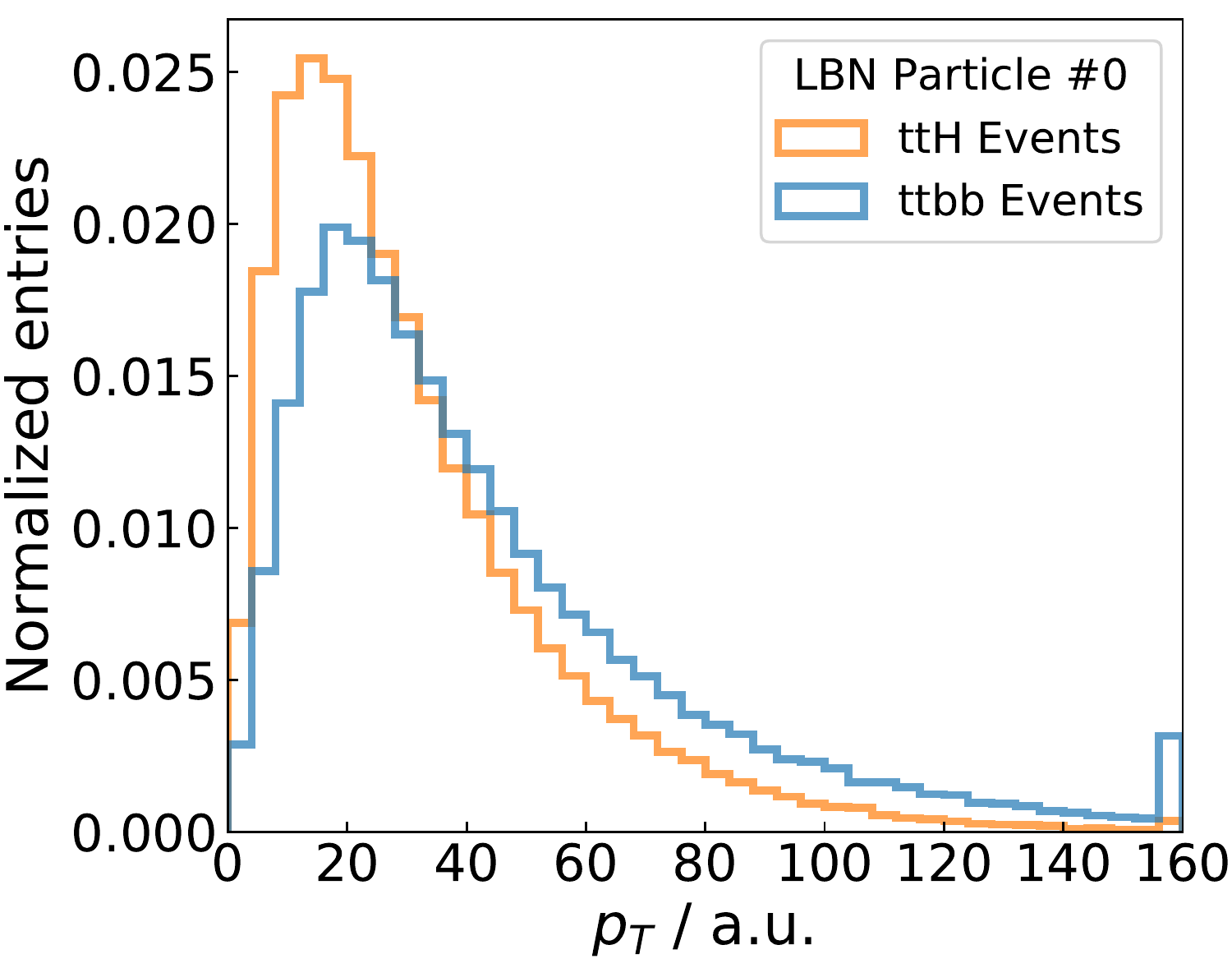}
        \caption{}
        \label{fig:slices_c}
    \end{subfigure}%
    \begin{subfigure}{.5\textwidth}
        \centering
        \includegraphics[width=0.8\textwidth]{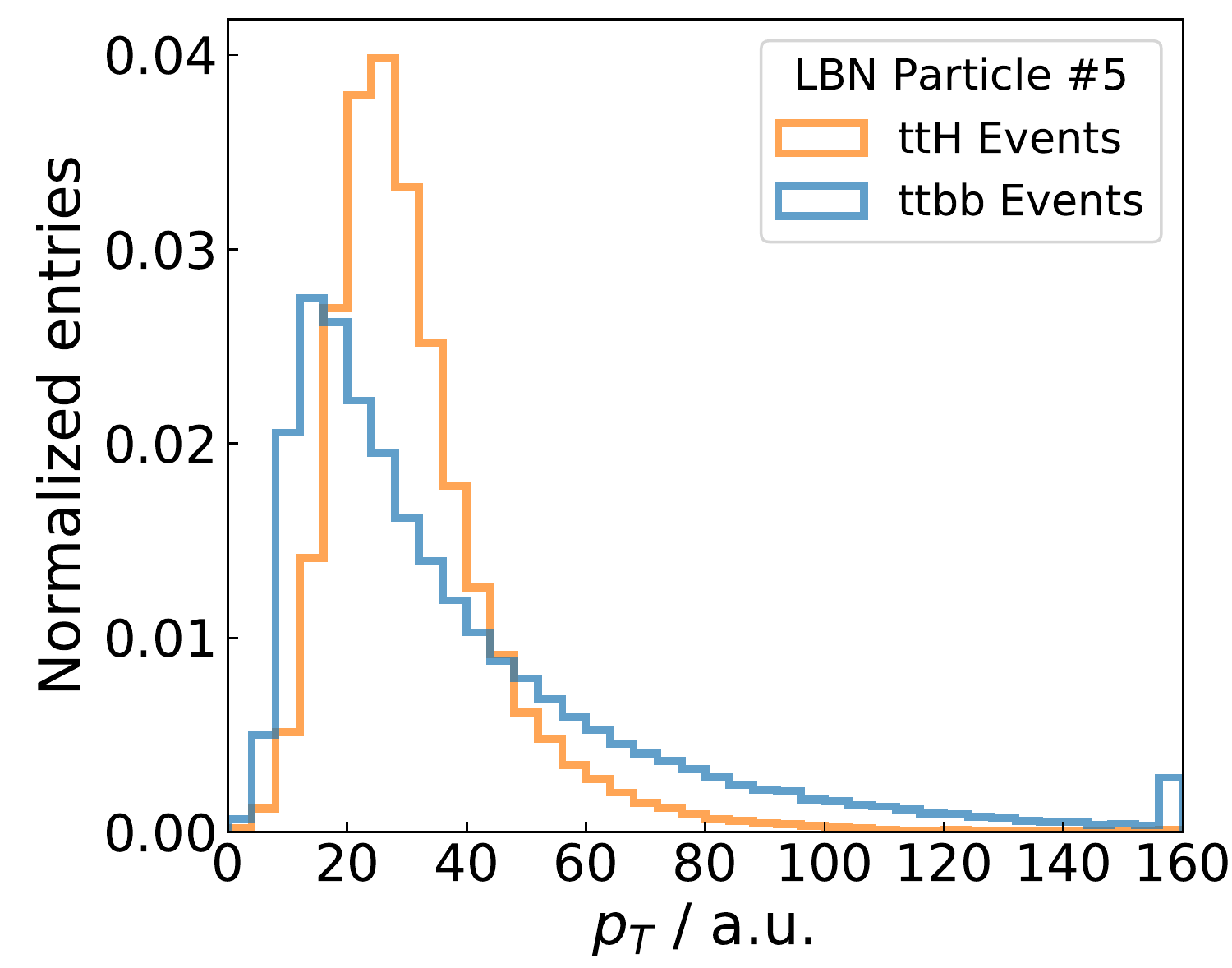}
        \caption{}
        \label{fig:slices_d}
    \end{subfigure}
    \begin{subfigure}{0.5\textwidth}
        \centering
        \includegraphics[width=0.8\textwidth]{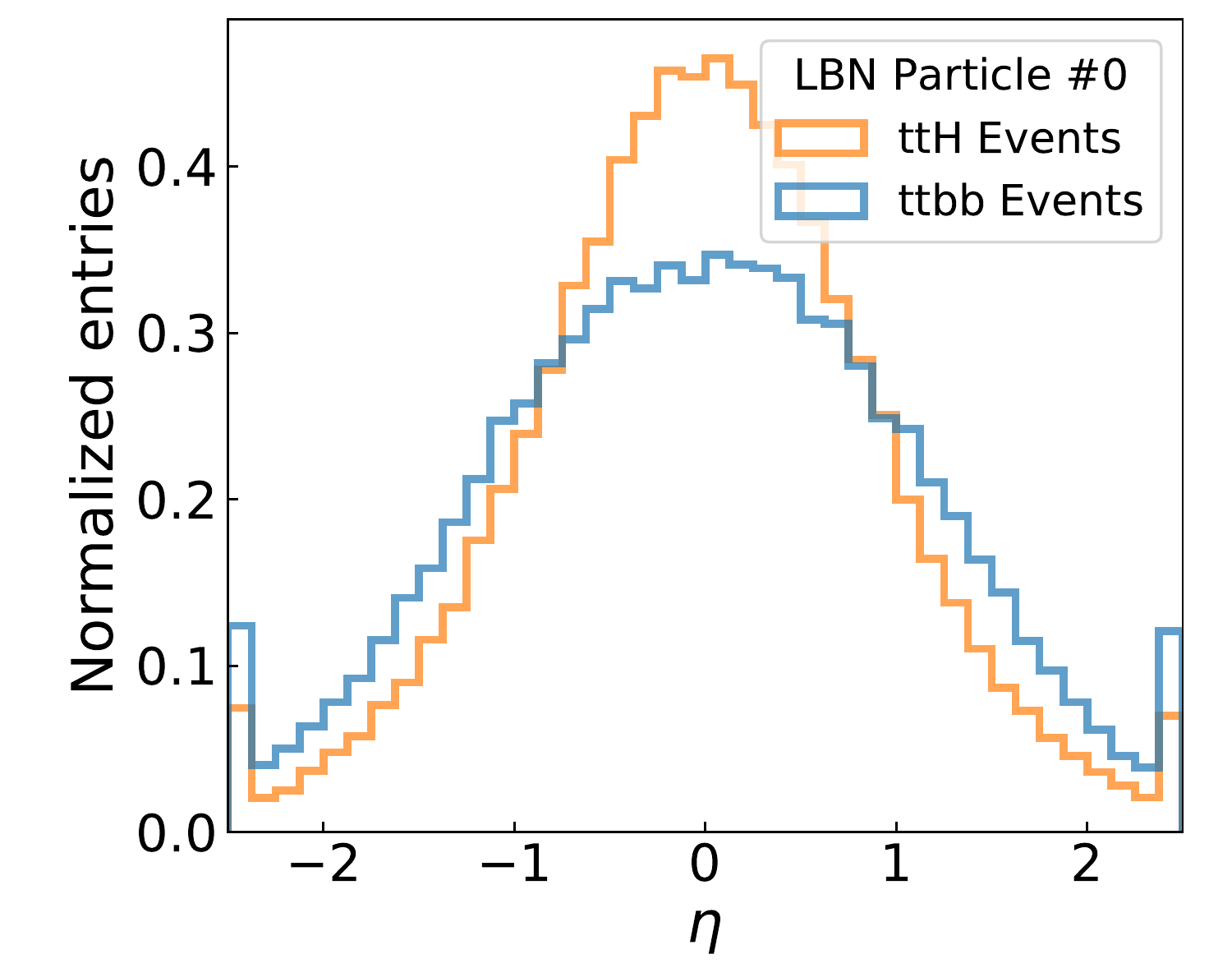}
        \caption{}
        \label{fig:slices_e}
    \end{subfigure}%
    \begin{subfigure}{.5\textwidth}
        \centering
        \includegraphics[width=0.8\textwidth]{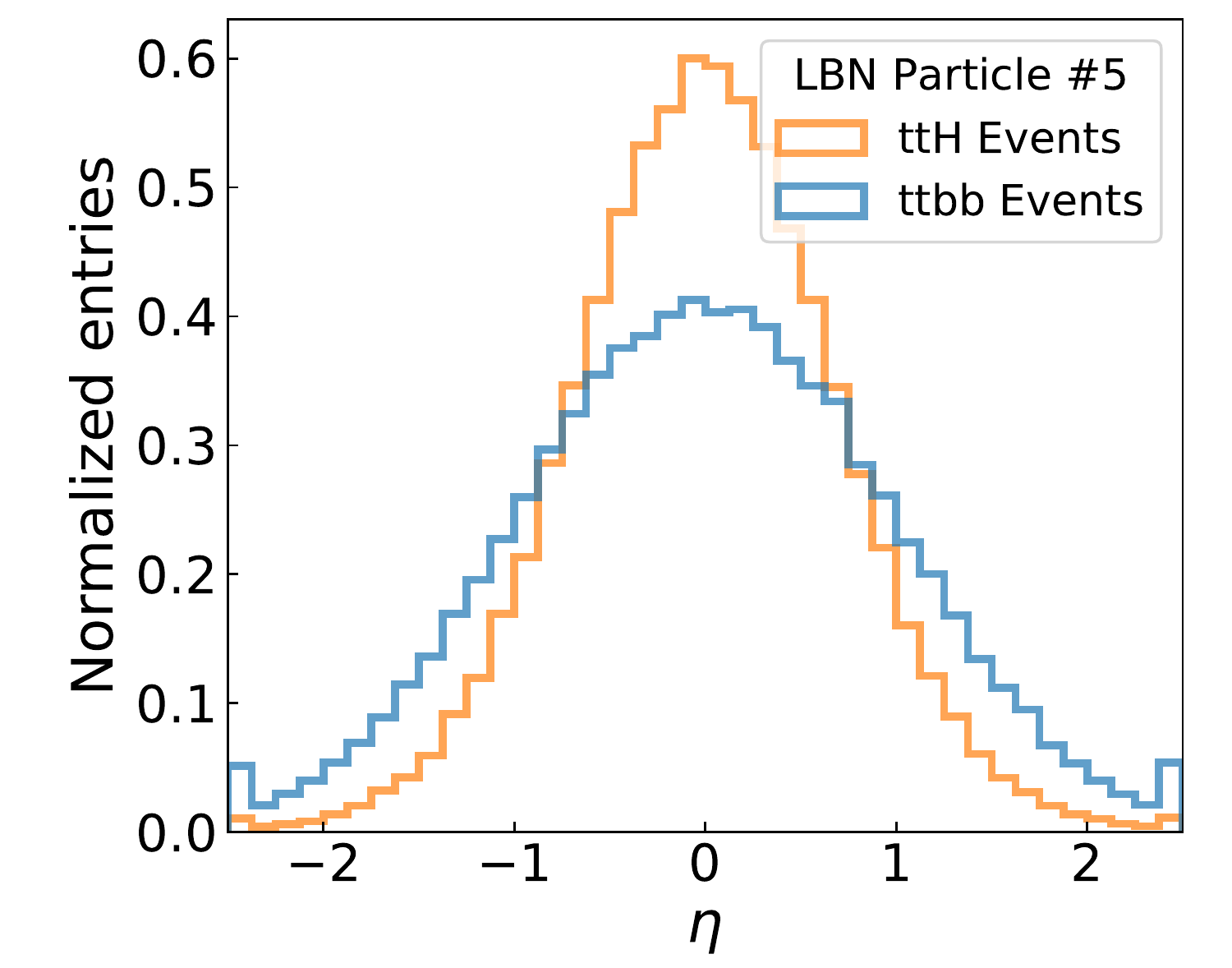}
        \caption{}
        \label{fig:slices_f}
    \end{subfigure}
    \caption{
        Example comparisons of kinematic distributions for $\ttH$ and $\ttbb$ processes.
        Figures a) and b) show the masses of the combined particles $0$ and $5$.
        Figures c)-f) also present their transverse momenta and pseudorapidities in the rest frames.
    }
    \label{fig:slices}
\end{figure}

For this Higgs-like configuration (combined particle $0$), we show mass distributions in \Fig{fig:slices_a}.
For $\ttH$, the invariant mass exhibits a narrow distribution around $m = 22$ a.u. reflecting the Higgs-like combined particle, while the $\ttbb$ background is widely distributed.
Note that because of the weighted four-vector combinations, arbitrary units are used, so that the Higgs boson mass is not stated in GeV.

\begin{figure}[h!tbp]
    \centering
    \begin{subfigure}{0.5\textwidth}
        \centering
        \includegraphics[width=0.9\textwidth]{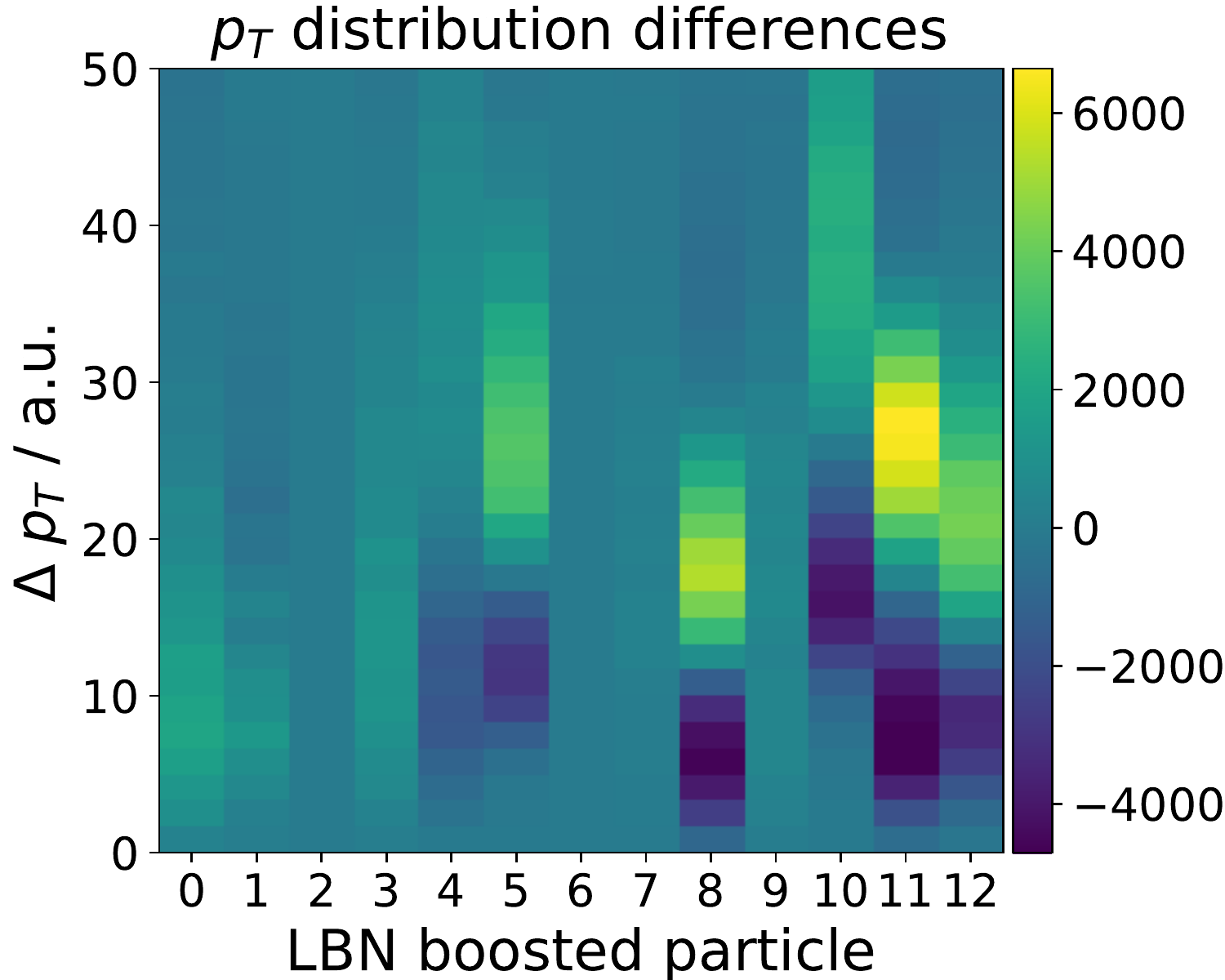}
        \caption{}
        \label{fig:gen_ttH_ttbb_a}
    \end{subfigure}%
    \begin{subfigure}{.5\textwidth}
        \centering
        \includegraphics[width=0.9\textwidth]{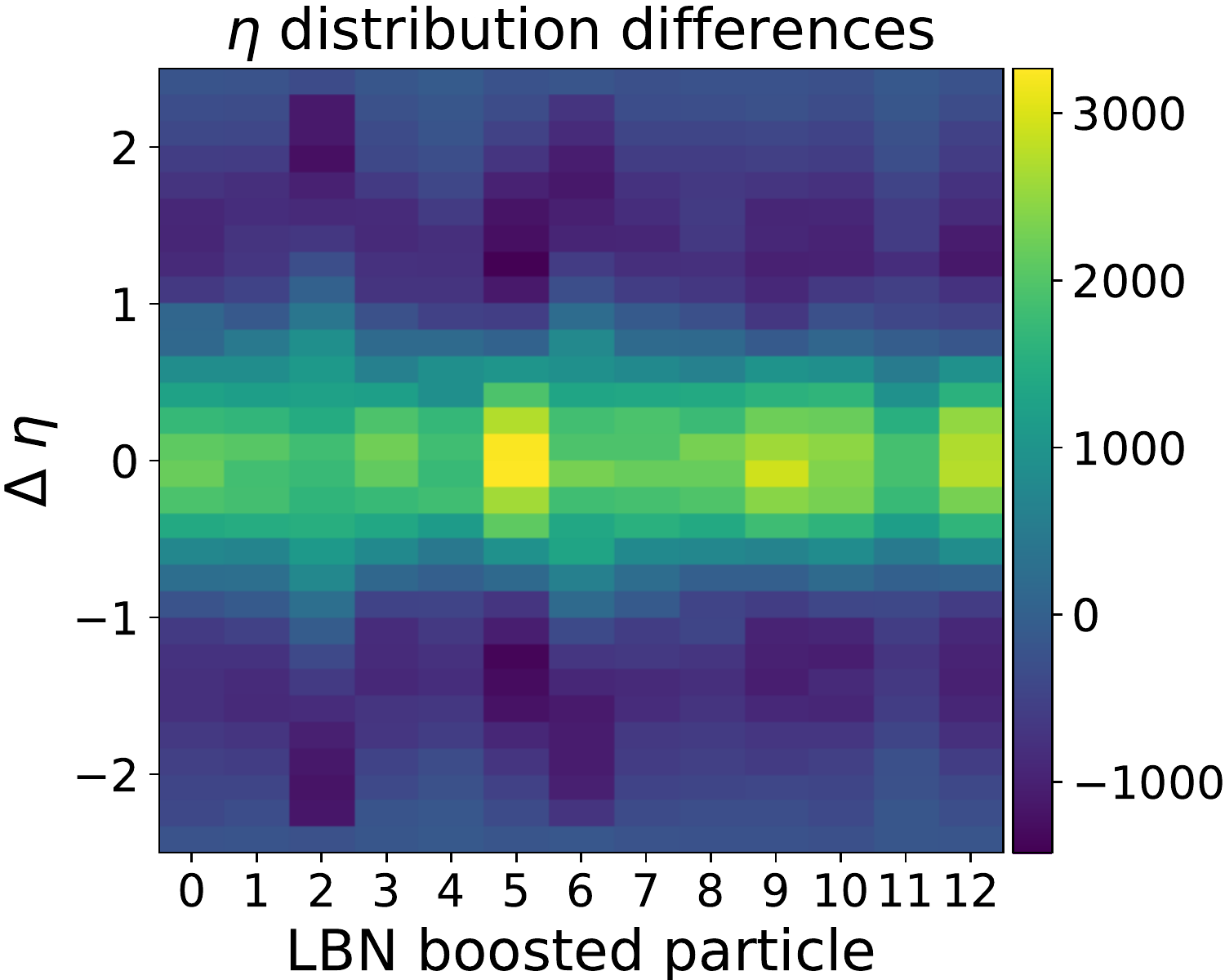}
        \caption{}
        \label{fig:gen_ttH_ttbb_b}
    \end{subfigure}
    \caption{
        Differences between $t\ttH$ and $\ttbb$ processes for combined particles in the rest frames of particle combinations, a) transverse momenta and b) pseudorapidities.
    }
    \label{fig:gen_ttH_ttbb}
\end{figure}

Analogously, we determine the transverse momentum $p_t$ and pseudorapidity $\eta$ distributions of the $13$ particles for signal and background events, and then compute their difference to visualize distinctions between $\ttH$ and $\ttbb$.
While the invariant mass distribution can also be formed without Lorentz transformation, both the $p_t$ and $\eta$ distributions are determined in the respective rest frames.
Clear differences in the distributions are apparent in \Fig{fig:gen_ttH_ttbb_a} for the transverse momenta of the combined particles $11, 8, 5, 10$ (in descending order of importance), while the combined particles $5, 9, 12, 10$ are most prominent for the pseudorapidities (\Fig{fig:gen_ttH_ttbb_b}).

As a selection of the many possible distributions, in \Fig{fig:slices} we show the invariant mass $m$, the transverse momentum $p_t$, and the pseudorapidity $\eta$ for two combined particles.
\Figs{fig:slices_a}, \ref{fig:slices_c}, and \ref{fig:slices_e} present the distributions of the combined Higgs-like particle $0$.
In addition to the prominent difference in the mass distributions, for $\ttH$ events its $p_t$ is smaller while its direction is more central in $\eta$ compared to $\ttbb$ events.

Furthermore, in \Figs{fig:slices_b}, \ref{fig:slices_d}, and \ref{fig:slices_f} we show combined particle $5$ because of its separation power in pseudorapidity and in transverse momentum (\Fig{fig:gen_ttH_ttbb}).
Combination $5$ is essentially determined by the bottom quark jet $\bi$ boosted into a rest frame of all other four-vectors (\Fig{fig:weight_gen_particles_normed}).
Here, the distribution of mass $m$ for $\ttH$ events is smaller, $p_t$ is larger and more defined, and $\eta$ is more central than for $\ttbb$ events.

In the example of the characteristic angular distribution of the charged lepton in top quark decays (\Fig{fig:lbn_example}, $\cos{\theta^*}$), two different reference systems are combined to determine the opening angle $\theta^*$ between the W boson in the top quark rest frame and the lepton in the W boson rest frame.
As mentioned above, the LBN is capable of calculating these complex angular correlations involving four-vectors evaluated in different rest frames.

\begin{figure}[h!tbp]
    \centering
    \begin{subfigure}{0.5\textwidth}
        \centering
        \includegraphics[width=0.8\textwidth]{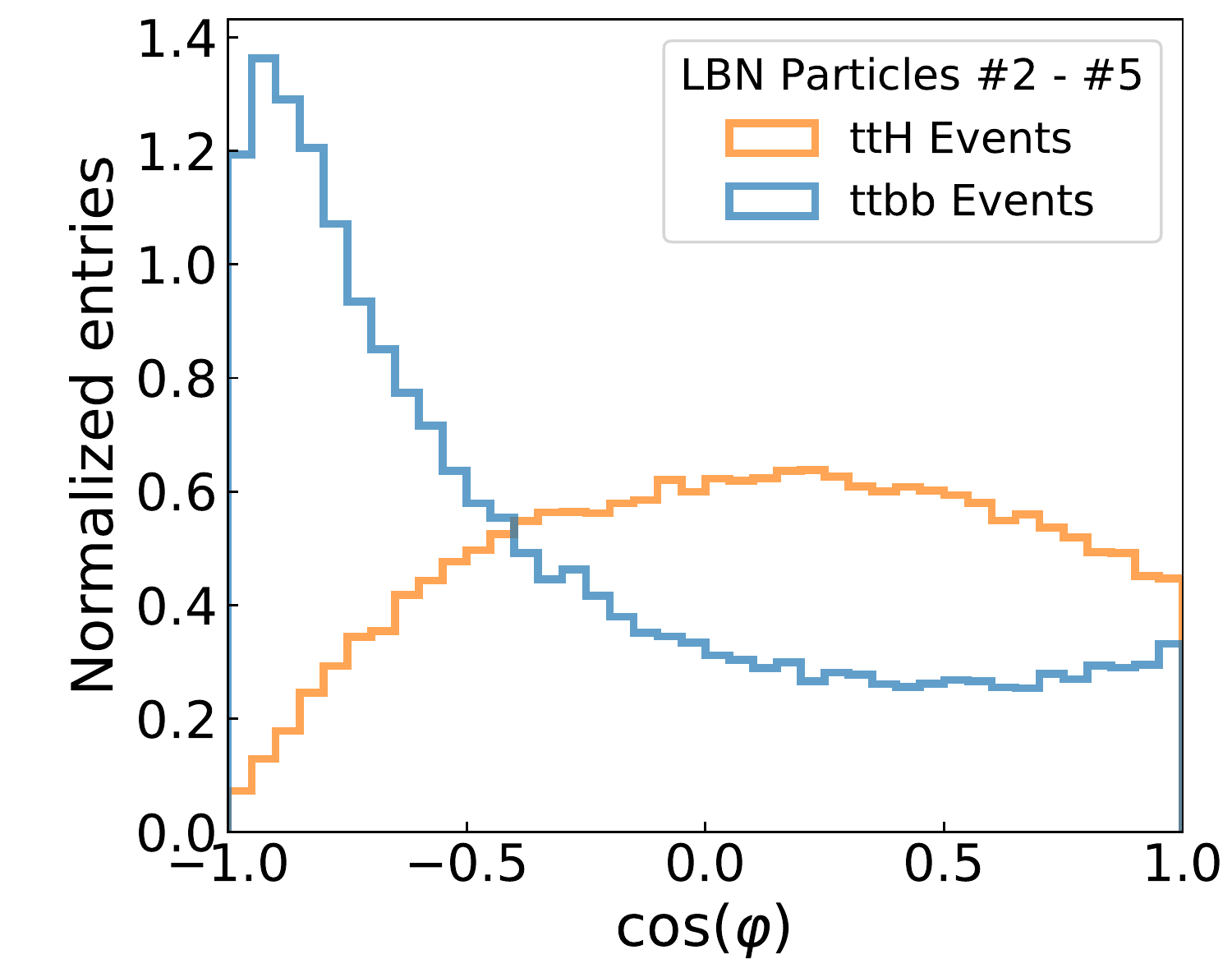}
        \caption{}
        \label{fig:cosangle_a}
    \end{subfigure}%
    \begin{subfigure}{.5\textwidth}
        \centering
        \includegraphics[width=0.8\textwidth]{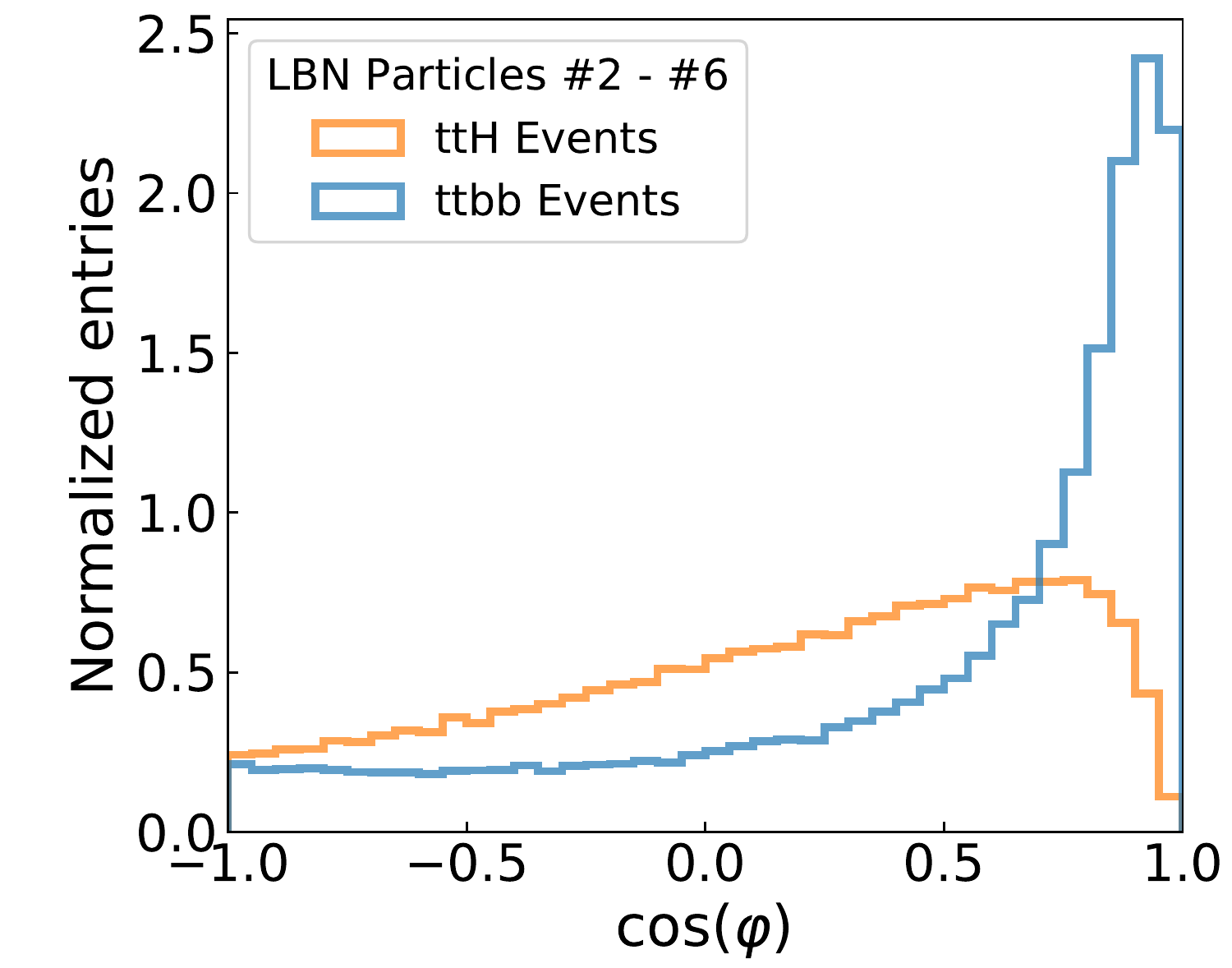}
        \caption{}
        \label{fig:cosangle_b}
    \end{subfigure}
    \caption{
        Cosine of the opening angles between combined particle $2$ ($\ttbar + \bi$) in its rest frame $\bii$ and a) combined particle $5$ ($\bi$) in its rest frame ($\ttbar + \bi + \bii$), and b) combined particle $6$ ($\ttbar + \bii$) in its rest frame ($\bi$).
    }
    \label{fig:cosangle}
\end{figure}

As an example, we select distributions for angular distances of combined particles $2$ and $5$ (\Fig{fig:weight_gen_particles_normed}).
Particle $2$ is a combination of all four-vectors boosted into the rest frame of the bottom quark jet $\bii$, whereas particle $5$ corresponds to the bottom quark jet $\bi$ boosted into a combination of all four-vectors.
\Fig{fig:cosangle_a} shows that, for $\ttbb$ background events, combined particles $2$ and $5$ predominantly propagate back-to-back, while the $\ttH$ signal distribution scatters broadly around $90$ deg.

The opposite holds true for the angles between combined particles $2$ and $6$ (\Fig{fig:weight_gen_particles_normed}).
Here, the bottom quark jet $\bi$, or alternatively $\bii$, serves as a rest frame for characterizing combinations of both top quarks together with the other bottom quark jet.
\Fig{fig:cosangle_b} shows that, for $\ttbb$, combined particles $2$ and $6$ preferably propagate in the same direction, whereas for $\ttH$ signal events this is less pronounced.

Overall, it can be concluded that the LBN network, when given only low-level variables as input, seems to autonomously build and transform particle combinations leading to extracted features that are suitable for the task of separating signal and background events.

\section{Conclusions}
\label{sec:conclusions}

The various physics processes from which the events of high-energy particle collisions originate lead to complex particle final states.
One reason for the complexity is the high energy that leads to Lorentz boosts of the particles and their decay products.
For an analysis task such as the identification of processes, we reconstruct aspects of the probability distributions from which the particles were generated.
How well probability distributions of individual physical processes are reflected is compiled in a series of characteristic variables.
To accomplish this, both the Minkowski metric for the correct calculation of energy-momentum conservation or invariant masses and the Lorentz transformation into the rest frames of parent particles are necessary.
So far, such characteristic variables have been engineered by physicists.

In this paper, we presented a general two-stage neural network architecture.
The first stage, the novel Lorentz Boost Network, contains at its core an efficient and fully vectorized implementation for performing Lorentz transformations.
The aim of this stage is to autonomously generate variables suitable for the characterization of collision events when using only particle four-vectors.
For this purpose, the input four-vectors are combined separately into composite particles and rest frames.
The composite particles are boosted into corresponding rest frames and a generic set of variables is extracted from the boosted particles.
The second stage of the network then uses these autonomously generated variables to solve a particular physics question.
Both the weights used to create the combinations of input four-vectors and the parameters of the second stage are learned together in a supervised training process.

To assess the performance of the LBN, we investigated a benchmark task of distinguishing $\ttH$ and $\ttbb$ processes.
We demonstrated the improved separation power of our model compared to domain-unspecific deep neural networks where the latter even used sophisticated high-level variables as input in addition to the four-vectors.
Furthermore, we developed visualizations to gain insight into the training process and discovered that the LBN learns to identify physics-motivated particle combinations from the data used for training.
Examples are top-quark-like combinations or the approximate compensation of the Lorentz boost of the center-of-mass system of the scattering process.

The LBN is a multipurpose method that uses Lorentz transformations to exploit and uncover structures in particle collision events.
It is part of an ongoing comparison of methods to identify boosted top quark jets and has already been successfully applied at the IML workshop challenge at CERN \cite{iml2018}.
The source code of our implementation in TensorFlow \cite{tensorflow} is publically available under BSD license \cite{lbncode}.

\begin{appendix}

\section*{Appendix}
\label{sec:appendix}

\subsection*{Network parameters}

The network parameters of the best performing networks are illustrated in Table~\ref{tab:network-parameters}.
\begin{table}[h!tbp]
    \centering
    \caption{
        All deep neural networks use a fully connected architecture with $n_\text{layers}$ and $n_\text{nodes}$.
        ELU is used as the activation function.
        The Adam optimizer \cite{kingma2014} is employed with decay parameters $\beta_1 = 0.9$, $\beta_2 = 0.999$, and a learning rate of $10^{-4}$.
        $L_2$ normalization is applied with a factor of $10^{-4}$.
        Batch normalization between layers was utilized in every configuration.
        The generic mappings in LBN feature extraction layer create $E, p_T, \eta, \phi, m$, and pairwise $\cos(\varphi)$.
    }
    \label{tab:network-parameters}
    \begin{tabular}{c|cc|cc|cc|cc}
        \toprule
        Network & \multicolumn{2}{c|}{\textbf{LBN+NN}} & \multicolumn{6}{c}{\textbf{DNN}}\\
        Variables & \multicolumn{2}{c|}{\textbf{low-level}} & \multicolumn{2}{c|}{\textbf{low-level}} & \multicolumn{2}{c|}{\textbf{high-level}} & \multicolumn{2}{c}{\textbf{combined}} \\
        Input Ordering & \textbf{gen.} & $\mathbf{p_T}$ & \textbf{gen.} & $\mathbf{p_T}$ & \textbf{gen.} & $\mathbf{p_T}$ & \textbf{gen.} & $\mathbf{p_T}$\\
        \midrule
        $M_\text{part.,rest fr.}$ & $13$ & $16$ & \verb|-| & \verb|-| & \verb|-| & \verb|-| & \verb|-| & \verb|-|\\
        $n_\text{layers}$ & $8$ & $8$ & $8$ & $8$ & $4$ & $4$ & $8$ & $6$\\
        $n_\text{nodes}$ & $1024$ & $1024$ & $1024$ & $1024$ & $512$ & $512$ & $1024$ & $1024$\\
        \bottomrule
    \end{tabular}
\end{table}

\subsection*{High-level variables}

The high-level variables employed in the training of the DNN benchmark comparison are inspired by \cite{Sirunyan:2018mvw}:
\begin{itemize}
    \item
    The event shape variables sphericity, transverse sphericity, aplanarity, centrality \cite{event_shape_variables}.

    \item
    The first five Fox-Wolfram moments \cite{fox_wolfram_moments}.

    \item
    The cosine of spatial angular difference $\theta^*$ between the charged lepton in the W boson rest frame and the W boson direction when boosted into the rest frame of its corresponding top quark.
    In the hadronic branch, the down-type quark is used owing to its increased spin analyzing power \cite{spin_analyzing_power}.

    \item
    The minimum, maximum and average of the distance in pseudorapidity $\eta$ and $\phi$ phase space $\Delta R$ of jet pairs.

    \item
    The minimum, maximum and average $|\Delta\eta|$ of jet pairs.

    \item
    The minimum and maximum of the distance in $\Delta R$ of jet-lepton pairs.

    \item
    The minimum, maximum and average $|\Delta\eta|$ of jet-lepton pairs.

    \item
    The sum of the transverse momenta of all jets.

    \item
    The transverse momentum and the mass of the jet pair with the smallest $\Delta R$.

    \item
    The transverse momentum and the mass of the jet pair whose combined mass is closest to the Higgs boson mass m$_H = 125$\,GeV \cite{CMS2012:higgs}.
\end{itemize}

\end{appendix}

\acknowledgments
We wish to thank Jonas Glombitza for his valuable comments on the manuscript.
We would also like to thank Jean-Roch Vlimant, Benjamin Fischer, Dennis Noll, and David Schmidt for a variety of fruitful discussions.
This work is supported by the Ministry of Innovation, Science and Research of the State of North Rhine-Westphalia, and by the Federal Ministry of Education and Research (BMBF).

\end{document}